\documentclass[american,5p,times,authoryear]{elsarticle}
\usepackage[T1]{fontenc}
\usepackage[latin9]{inputenc}
\usepackage{xcolor}
\usepackage{array}
\usepackage{wrapfig}
\usepackage{booktabs}
\usepackage{textcomp}
\usepackage{multirow}
\usepackage{amsmath}
\usepackage{amssymb}

\usepackage{setspace}
\usepackage{subscript}
\makeatletter

\providecommand{\tabularnewline}{\\}

\journal{Energy Economics (accepted for publication in Energy Economics: 06 Oct 2018)}

\usepackage{amsmath,epsfig,amssymb,graphicx}
\usepackage{bbm}
\usepackage{setspace,exscale,latexsym,srcltx}
\usepackage{longtable,pdflscape}
\usepackage[scanall]{psfrag}
\usepackage{rotating}
\usepackage[nottoc]{tocbibind}
\usepackage{dsfont}

\usepackage{tikz}
\usepackage{subcaption}

\usepackage{natbib}
\usepackage{breakurl} 

\usepackage{url}

\usepackage{rotating}
\usepackage{tocloft}
\usepackage{tikz}
\usepackage{color}
\usepackage{tocbibind}
\usepackage{changepage}

\usepackage[labelfont=bf]{caption}

\usepackage[english]{babel}
\usepackage{wrapfig}
\DeclareMathOperator*{\argmin}{arg\,min}

\usepackage{booktabs}

\usepackage{har2nat}

\usepackage{fancyhdr}	
\allowdisplaybreaks

\makeatother

\usepackage{babel}
\begin{document}

\begin{frontmatter}{}

\title{The value of forecasts: Quantifying the economic gains of accurate
quarter-hourly electricity price forecasts}

\author{Christopher Kath\fnref{fn1}\textsuperscript{a,{*},}}

\ead{christopher.kath@rwe.com}

\author{Florian Ziel\textsuperscript{b}}
\begin{abstract}
We propose a multivariate elastic net regression forecast model for
German quarter-hourly electricity spot markets. While the literature
is diverse on day-ahead prediction approaches, both the intraday continuous
and intraday call-auction prices have not been studied intensively
with a clear focus on predictive power. Besides electricity price
forecasting, we check for the impact of early day-ahead (DA) EXAA
prices on intraday forecasts. Another novelty of this paper is the
complementary discussion of economic benefits. A precise estimation
is worthless if it cannot be utilized. We elaborate possible trading
decisions based upon our forecasting scheme and analyze their monetary
effects. We find that even simple electricity trading strategies can
lead to substantial economic impact if combined with a decent forecasting
technique.
\end{abstract}

\address{\textsuperscript{a}RWE Supply \& Trading GmbH, Altenessenerstr.27,
45141 Essen }

\address{\textsuperscript{b}University of Duisburg-Essen, House of Energy Markets and Finance}

\ead{florian.ziel@uni-due.de}

\fntext[fn1]{The findings, interpretations and conclusions expressed hereinafter are those of the author and do not necessarily reflect the views of RWE Supply \& Trading GmbH.}
\begin{keyword}
forecasting, portfolio analysis, elastic net regression, Markowitz
portfolio, quarter-hourly spot prices, electricity price forecast\cortext[cor1]{Corresponding author}
\end{keyword}

\end{frontmatter}{}

\section{Introduction}

\label{sec:1}Germany is an outstanding example of massive renewable
integration within the European energy market. Politically induced,
renewable generation capacities were expanded and their marketing
subsidized. This not only affected the German day-ahead bid-stack
but also caused exchanges and market participants likewise to set
the focus on quarter-hourly (QH) considerations for their optimization
procedures due to the increasing residual volumes after hourly day-ahead
bidding. For more information on the described renewables impact,
the interested reader might refer to \citet{hirth2013market}; \citet{paraschiv2014impact};
\citet{ketterer2014impact}; \citet{wurzburg2013renewable}. As a
result of this ongoing trend, marketplaces have adapted their products
so that the German market features another unique characteristic\textcolor{violet}{.}
While other countries such as the Netherlands or Belgium do not offer
any possibility to trade QH products at the time of the writing of
this paper, Germany has three independent exchanges that allow trading
on an early day-ahead basis up to half an hour before physical delivery.
The opportunity to enter QH trades started in December 2011 with the
first 15-minute contracts in continuous intraday markets and was consequently
expanded in September and December 2014 by EXAA quarter-hourly day-ahead
products and the EPEX intraday call auction. A more thorough discussion
of the German spot market is provided by \citet{viehmann2017state}.
\\
\hspace*{0.5cm}Unfortunately, academic attention is only recently
focused on lower time intervals. Discussions of quarter-hourly German
spot markets are rare. A good starting point is provided by \citet{kiesel2017econometric}
who discuss the econometric characteristics of quarter-hourly EPEX
intraday (ID) time series and provide an analytical model approach.
\citet{markle2018contract} evaluate market impacts of the introduction
of 15-minute contracts and report price reductions in correlated hourly
spot markets. However, the current literature lacks a decent discussion
of forecasting QH prices. Quarter-hourly trading appears to be crucial,
but there is no particular forecasting model available. This statement
equally counts for QH auctions as well as continuous intraday trading.
We aim to fill this gap by providing precise price estimations for
both of these markets. To achieve this, we will consider the most
current input factors in German spot trading together with the status
quo in forecasting techniques. \\
\hspace*{0.5cm}Another aspect that must not be ignored in this context
is the economic effect of an estimation scheme. On the one hand, many
forecasting models exist, at least for hourly day-ahead applications
(see \citet{weron2014electricity} for a broader discussion), on the
other hand, the majority of these limit their scope to the evaluation
of accuracy but neglect the aspect of economic benefits. Even the
most accurate prediction has no practical value if done in a market
or at a point in time where no possibility of a utilization exists.
Therefore, our second contribution shall be a quantification of attainable
gains through precise forecasts in QH spot markets.\\
\hspace*{0.5cm}The rest of this paper is divided into the following
sub-sections: Section \color{blue} \ref{sec:2} \color{black} introduces
available German QH spot markets and highlights their peculiarities,
followed by section \color{blue} \ref{sec:3} \color{black} discussing
the connected forecast methodology. This comprises the model input
parameters, necessary data transformations and the overall estimation
algorithm. Section \color{blue} \ref{sec:4} \color{black} addresses
the forecast performance in our empirical study and the associated
economic effects of our price predictions followed by a conclusion
and a short outlook on further expansions in section \color{blue}\ref{sec:outlook}\color{black}. 

\section{Quarter-hourly trading and its relevance in Germany}

\label{sec:2}
\begin{figure*}[!t]
\centering{}\includegraphics[scale=0.6]{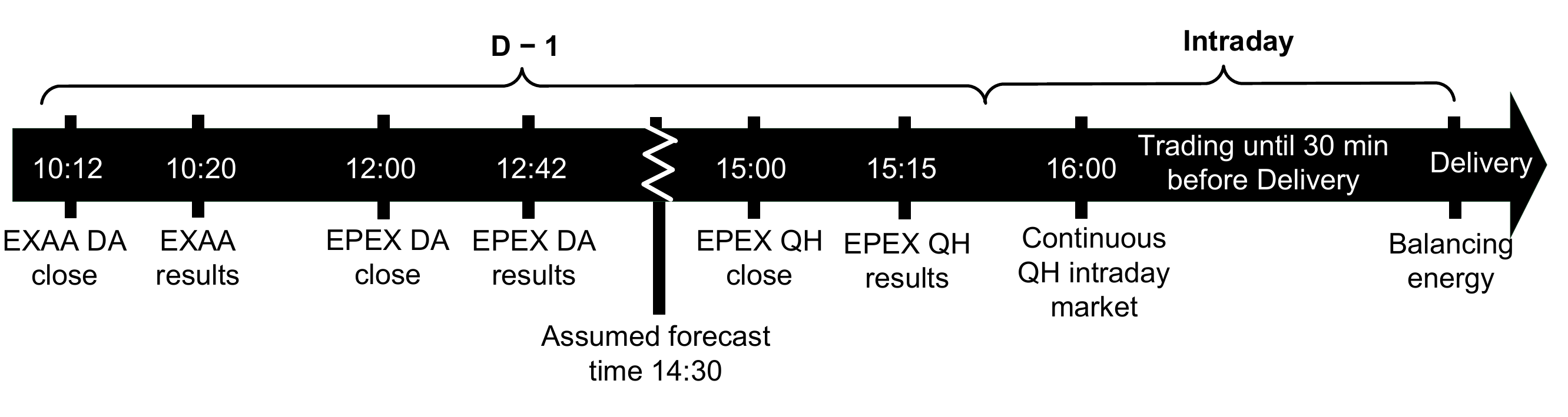}\caption{Spot trading timeline and its connected trading venues. All mentioned
deadlines are assuming usual circumstances with no delayed results,
technical problems and a market clearing price determined in the first
run of the underlying algorithm. Please note that the intraday lead
time of 30 minutes applies for cross-grid trades and local trades
within the same grid area are allowed until 5 minutes before delivery.}
\end{figure*}
Germany offers a wide variety of possible trading venues for market
participants. Other countries usually exhibit a day-ahead spot exchange
and continuous intraday trading platforms. These are also to be found
for the four German grid areas, but besides them, there are two other
auctions, as depicted in Figure 1. Spot trading ideally starts with
the EXAA (Energy Exchange Austria) at 10:12am for final bid submission.
Only 8 minutes later, the EXAA publishes the first day-ahead exchange
traded quotation for the German delivery area. Although an Austrian
exchange, EXAA results can easily be delivered into German market
areas. However, we must acknowledge that this situation could be of
temporary character with ongoing talks about splitting the German-Austrian
bidding zone.\footnote{As per June 2018, when this paper was finalized, implementation of
a market split had not been achieved. Therefore, possible effects
of a German-Austrian split are uncertain and ignored in the following.} As a result, EXAA volumes might only be transferred with explicitly
sold cross-border capacities or are implicitly regarded by exchange
auctions. One feature only available with EXAA is post-trading. The
exchange platform allows for a second bidding round with known prices
to market a surplus on either the buy or sell side. EXAA trading only
occurs on non-holiday weekdays. All weekend or holiday prices are
determined in advance on the last weekday before the holiday or weekend.
Therefore, we already have a QH indication for delivery date Sunday
on Friday, for instance. The next and\begin{wraptable}{o}{0.66\columnwidth}%
\begin{centering}
\begin{tabular}{>{\centering}p{2.3cm}ccc}
\hline 
{\footnotesize{}Exchange} & \multicolumn{3}{c}{{\footnotesize{}traded volume {[}TWh{]}}}\tabularnewline
 & {\footnotesize{}2015} & {\footnotesize{}2016} & {\footnotesize{}2017}\tabularnewline
\hline 
{\footnotesize{}EPEX DA auction} & {\footnotesize{}264} & {\footnotesize{}235} & {\footnotesize{}233}\tabularnewline
{\footnotesize{}EXAA DA auction} & {\footnotesize{}8.2} & {\footnotesize{}8.0} & {\footnotesize{}5.4}\tabularnewline
{\footnotesize{}EPEX QH auction} & {\footnotesize{}3.9} & {\footnotesize{}4.6} & {\footnotesize{}5.2}\tabularnewline
{\footnotesize{}EPEX QH ID } & {\footnotesize{}3.9} & {\footnotesize{}3.6} & {\footnotesize{}4.9}\tabularnewline
\hline 
\end{tabular}
\par\end{centering}
\caption{Yearly volumes of hourly and quarter-hourly German spot exchanges.
All intraday figures only entail data on Germany, while the day-ahead
auction includes Austria and Luxembourg.}
\end{wraptable}%
 presumably most important trading opportunity is provided by the
German EPEX day-ahead (DA) auction. A single bidding round with results
available at 12:42am marks the primarily traded market quotation in
the day-ahead market. At the time of writing, the term 'EPEX day-ahead'
correctly specifies the German hourly day-ahead exchange. Still, the
other exchanges, EXAA and Nord Pool Spot, are expanding their activities
to the German day-ahead market. It is planned to unbundle the pricing
algorithm from EPEX such that three independent exchanges offer access
to the price that is hereinafter referred to as 'EPEX day-ahead'.
We stick to that notation to be in line with other literature and
due to the fact that these changes are planned but have not been implemented
yet. \\
\hspace*{0.5cm}Due to rising renewables infeed and the necessity
to balance quarter-hourly deviations, EPEX launched a second auction
for quarter-hours in December 2014. Strictly chronologically speaking
it takes place day-ahead, nevertheless it is referred to as an intraday
call auction because the day-ahead market window ends at d-1, 14:30pm
for grid operators, as depicted by the white lines in Figure 1. While
all prior marketplaces allow entering a single round of bids determining
the price level in a closed-form auction, our last trading opportunity,
the EPEX intraday market, is a continuous one that is tradable up
to 30 minutes before delivery. This lead time was changed per July
2015 from 45 to 30 minutes. We will consider the volume weighted average
price (VWAP) of all transactions for the specific delivery quarter-hour
since continuous trading activities are difficult to quantify otherwise.
Last but not least, all open positions will be settled by the grid
operators in the course of balancing energy at the grid area independent
imbalance tariff (reBAP). Since it is strictly forbidden by regulators
to enter imbalance positions intentionally, this market is not a trading
alternative and is just mentioned for the sake of comparability.\\
\hspace*{0.5cm}Table 1 hints at the relevance of the different exchanges.
The allocation of volumes points towards the immense importance of
the hourly EPEX DA auction. It outruns the QH trading venues by far.
This phenomenon might be explained by their purposes. As a result
of missing liquidity, market players are more likely trading residual
positions in QH markets. The majority, i.e., the hourly demand and
generation will be bid in the day-ahead exchange for which reason
QH liquidity only accounts for 2\% of the DA liquidity. Unfortunately,
EXAA volumes are reported in an aggregated form without any separation
into hourly or quarter-hourly amounts. Hence, the mentioned trading
volumes only allow for a rough evaluation of importance. The low volumes
suggest that the EPEX markets are more momentous when German spot
trading is concerned. Whenever liquidity is limited, this could elicit
high volatility and price spikes. To detect such occurrences, we have
plotted the price series in Figure 2. Both the QH auction and the
ongoing QH intraday trading can be highly volatile with prices under
0€/MWh or above 100€/MWh. While, in general, both time series appear
to follow similar trends, the intraday equivalent seems to feature
more spikes. However, this effect is not predominant. The overall
picture reflects two resemblant price quotations.
\begin{figure*}
\begin{centering}
\includegraphics[scale=0.595]{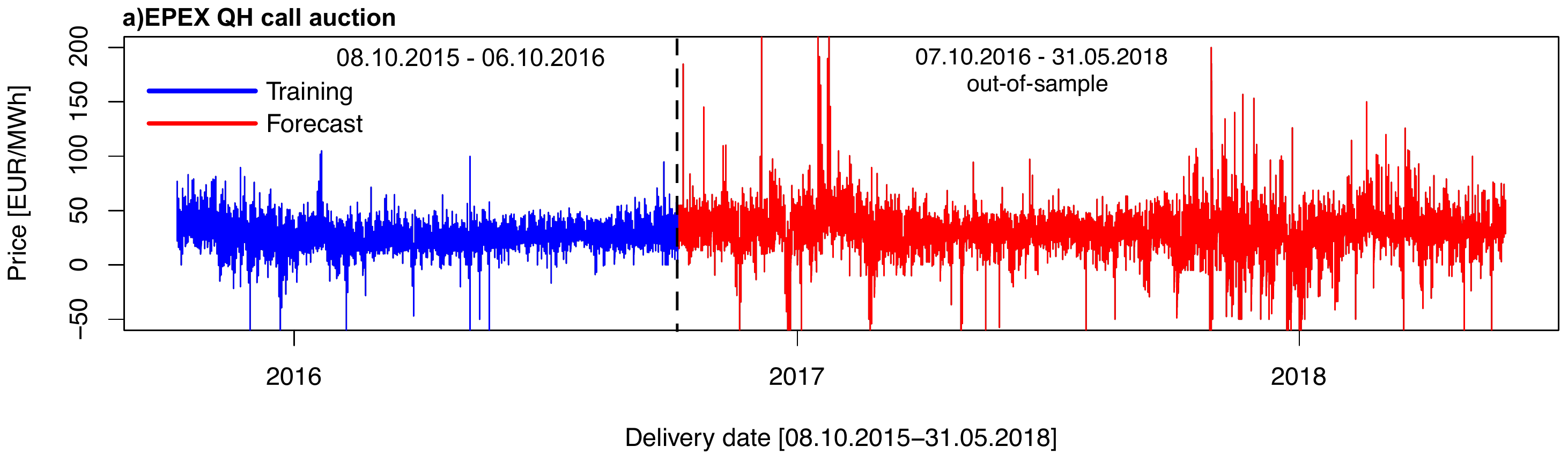}
\par\end{centering}
\centering{}\includegraphics[scale=0.593]{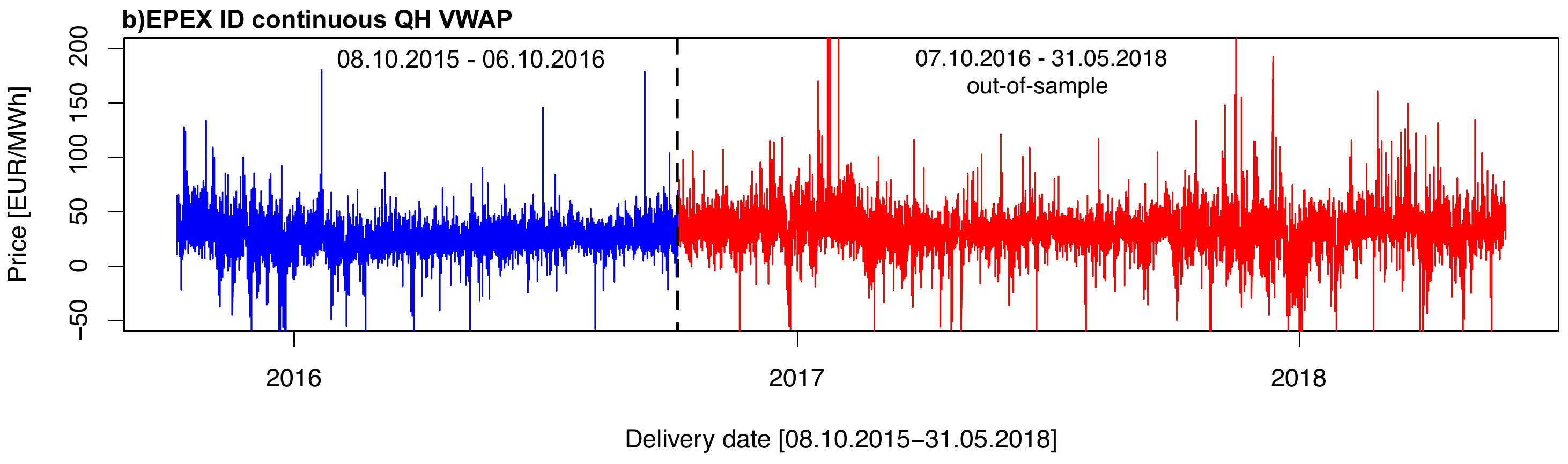}\caption{Price plot of the EPEX intraday QH auction and EPEX intraday continuous
price regime separated into training and forecast sections. The blue
partition marks the initial training and parameterization period that
is consequently shifted with each iteration of the rolling estimation.
The red line depicts the out-of-sample data that our prediction models
try to forecast.}
\end{figure*}

\section{Forecast methodology\label{sec:3}}

\subsection{Data transformation and input parameters}

The price plots reveal price spikes and the occurrence of negative
prices. This is not a general problem but would usually require either
an explicit modeling of spikes by means of a price spike component,
a spike-robust model or a transformation to stabilize the variance
of the time series (\citet{uniejewski2018variance}). We have decided
on the latter as we do not want to give up the feature selection abilities
of our models discussed later. Once transformed, one can use a wider
set of algorithms without taking greater care of price spikes. We
firstly transform and then inverse the data such that the output of
our models still appears in a realistic format. The transformation
mainly supports the algorithms by providing a more stable variance
but does not change any crucial information. \\
\hspace*{0.5cm}A usual way to transform price series is the logarithm.
While a simple logarithmic transformation works in many different
scenarios, our time series with negative values necessitates a transformation
method that can handle negative values. We stick to current literature
findings to identify the best transformation for our needs. In a large
empirical study, \citet{uniejewski2018variance} report superior RMSE-related
performance for a newly proposed transformation called 'mlog', which
we utilize for this paper. The authors especially propose the transformation
for the spike sensitive measure RMSE (root-mean-square-error)\footnote{Please refer to section 4.1 for the mathematical formulation of RMSE.}
which makes sense to apply to highly volatile time series such as
our intraday one. The mlog transformation showed constant results
across all markets, which is why we decided to use it for our time
series and markets. Before its actual processing, the data requires
normalization. Hence, the original time series $x_{qh,t}$ is adjusted
to $z_{qh,t}=\frac{1}{MAD}(x_{qh,t}-median)$ in which MAD describes
the median absolute deviation (MAD). Both MAD and median are calculated
for $x_{qh,t}$ over the entire period. We purposely introduce a neutral
time series notation $x_{qh,t}$ since the transformation procedure
is not only executed on prices but on also other external factors
like load or wind. Once the data is normalized, its transformation
$y_{qh,t}$ is given by (taken from \citet{uniejewski2018variance})
\begin{equation}
y_{qh,t}=sgn(z_{qh,t})\left[\mathrm{log}(\left|z_{qh,t}\right|+\frac{1}{c})+\mathrm{log}(c)\right],
\end{equation}
and its inverse function 
\begin{equation}
z_{qh,t}=sgn(y_{qh,t})\left[e^{\left|z_{qh,t}\right|-log(c)}-\frac{1}{c}\right],
\end{equation}
with $c=\frac{1}{3}$. This parameter was likewise used by \citet{uniejewski2018variance}
and yielded good results across several markets.\\
\hspace*{0.5cm}The time series is a quarter-hourly one which renders
a slight transformation necessary. Daylight saving time causes one
duplicate hour as well as a missing value. We follow \citet{weron2007modeling}
and average the duplicative hour. Its omitted equivalent is calculated
using multiple imputations as presented in \citet{buuren2011mice}
so that every day in the empirical test consists of 96 QHs. We also
apply this approach to all other gaps in the time series. Apart from
that, no more pre-processing is carried out. We neglect all outlier
effects in our estimation scenario and leave extreme values untouched.
Our empirical sample ranges from 08.10.2015 to 31.05.2018. Instructions
on how to obtain the different data series are provided in Table 2.
A solely autoregressive approach is not desirable as many papers suggest
the influence that external factors have. \\
\begin{figure*}[!t]
\centering{}\includegraphics[scale=0.65]{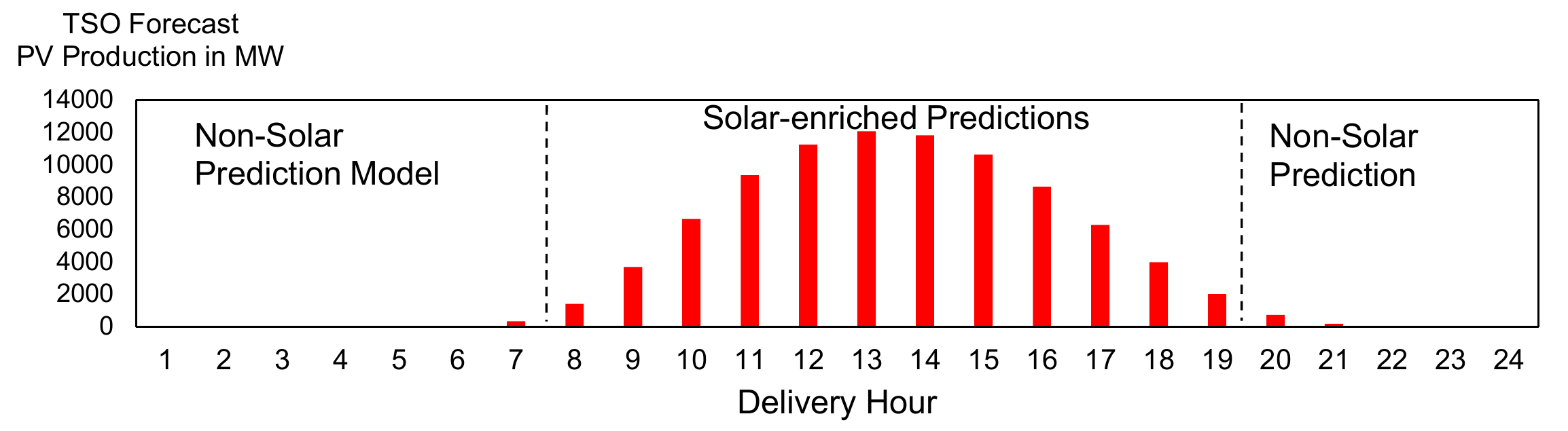}\caption{Averaged photovoltaics (PV) production forecast for Germany based
on the publicly available TSO PV forecast. Please note that we have
considered our entire time series from 01.10.2015 to 31.05.2018 and
averaged across all 24 delivery hours in order to analyze any day
and night effects. }
\end{figure*}
\hspace*{0.5cm}We aim to keep the model simple and easily reproducible
and only consider the most common publicly available external parameters
like the quarter-hourly ENTSO-E load forecast (e.g. used in \citet{kiesel2017econometric})
or wind power reported by the EEX transparency platform (see \citet{pape2016fundamentals,aid2016optimal,garnier2015balancing}
for models that include wind infeed). The two input factors are fundamentally
driven and might feature ramping effects. For instance, morning times
when industrial shifts begin and people are waking up cause the grid
load to quickly increase, whereas its level is more likely to be stable
around noon. We embrace these effects for wind power production and
load by regarding not only the load or wind infeed forecast for a
specific hour but also the forecast from one hour previous. Strong
differences between the two values might indicate ramping effects
and can contain valuable information for our prediction model. Connected
to these inputs is the concern over hourly data. Some prices and the
wind data are present in hourly formats only. They are transformed
rather modestly by assuming the hourly values for every quarter-hour
without any further processing. Since we do not know anything about
the quarter-hourly allocations, this seems to be the most unbiased
way to capture these effects. As for wind, one might also find quarter-hourly
forecasts by professional providers. We have deliberately chosen the
hourly TSO data to ensure high reproducibility, but need to concede
that designated vendor data increases forecast accuracy since it provides
more accurate QH weather data. \\
\hspace*{0.5cm}Speaking of weather data, one must not forget the
other crucial component of the German fuel mix: Photovoltaics (PV)
generation. A clear sign of its importance is that even the exchange
itself mentions PV infeed as one of the major reasons for the introduction
of the QH auction in 2013 (see \citet{epex15min} for the press release).
\citet{markle2018contract} support this assessment of importance
by stating that QH trading is mostly driven by PV ramp-ups or -downs,
i.e., times when PV production quickly increases or decreases. However,
a forecaster needs to be careful with PV data. During the night, the
time series features a constant zero due to no production which might
cause problems with prediction models. Figure 3 illustrates how this
effect is allocated over an entire day. The averaged PV production
only starts to remarkably differ from zero in a time frame between
hours 8 and 19. We have made the expert decision to add PV production
data to all QH prediction models from quarter-hours 29 to 76 and ignore
PV entirely in case of all other quarter-hours. We also want to capture
ramping effects as in wind and load forecasts and consider the official
TSO PV infeed forecast for the relevant hour together with its equivalent
prediction one period before. Hence, our prediction approach accounts
for ramp-ups or ramp-downs in PV production. \\
\hspace*{0.5cm}Figure 1 is not strictly limited to quarter-hourly
markets, but if we do so, three trading opportunities remain: the
EPEX QH auction, continuous intraday trading and the EXAA auction
which publishes results at 10:20am a day ahead. Therefore, the first
quarter-hourly price information is delivered by EXAA prices. Its
information might be incorporated into a forecasting scheme for the
EPEX markets (see \citet{ziel2017modeling} for this thought).
\begin{table*}
\begin{centering}
\begin{tabular}{>{\centering}p{2cm}>{\centering}p{2cm}>{\centering}p{6cm}>{\centering}p{4cm}>{\centering}p{2cm}}
\hline 
{\footnotesize{}Determinant} & {\footnotesize{}Unit/granularity} & {\footnotesize{}Description} & {\footnotesize{}Data source} & {\footnotesize{}Transformation}\tabularnewline
\hline 
{\footnotesize{}EPEX day-ahead auction price} & {\footnotesize{}EUR/MWh, hourly} & {\footnotesize{}Market clearing price of the EPEX day-ahead auction,
physical delivery into German or Austrian grid possible} & {\footnotesize{}European Power Exchange (EPEX), \href{https://www.epexspot.com/en/}{https://www.epexspot.com/en/}} & {\footnotesize{}mlog, hourly value for all QHs}\tabularnewline
{\footnotesize{}EPEX intraday auction price} & {\footnotesize{}EUR/MWh, quarter-hourly} & {\footnotesize{}Market clearing price of the EPEX intraday auction,
physical delivery into German grid } & {\footnotesize{}European Power Exchange (EPEX), \href{https://www.epexspot.com/en/}{https://www.epexspot.com/en/}} & {\footnotesize{}mlog}\tabularnewline
{\footnotesize{}EPEX intraday VWAP} & {\footnotesize{}EUR/MWh, quarter-hourly} & {\footnotesize{}Volume weighted average of all transactions for specific
QH, physical delivery into German grid} & {\footnotesize{}European Power Exchange (EPEX), \href{https://www.epexspot.com/en/}{https://www.epexspot.com/en/}} & {\footnotesize{}mlog}\tabularnewline
{\footnotesize{}EXAA day-ahead auction price} & {\footnotesize{}EUR/MWh, quarter-hourly} & {\footnotesize{}Market clearing price of the EXAA day-ahead auction,
physical delivery into German and Austrian grid possible} & {\footnotesize{}Energy Exchange Austria (EXAA), \href{http://www.exaa.at/en}{http://www.exaa.at/en}} & {\footnotesize{}mlog}\tabularnewline
{\footnotesize{}ENTSO-E }{\footnotesize \par}

{\footnotesize{}load forecast} & {\footnotesize{}MW, quarter-hourly} & {\footnotesize{}Vertical system load for bidding zone Germany/Austria,
published around 10:00 d-1} & {\footnotesize{}European Network of Transmission System Operators
(ENTSO-E), \href{https://transparency.entsoe.eu/}{https://transparency.entsoe.eu/}} & {\footnotesize{}mlog}\tabularnewline
{\footnotesize{}TSO PV forecast} & {\footnotesize{}MW, hourly} & {\footnotesize{}Photovoltaics infeed forecast for Germany published
by transmission system operators (TSO) at 8:00 d-1} & {\footnotesize{}European Energy Exchange (EEX),}{\footnotesize \par}

{\footnotesize{}\href{https://www.eex-transparency.com/}{https://www.eex-transparency.com/}} & {\footnotesize{} mlog, hourly value for all QHs}\tabularnewline
{\footnotesize{}TSO wind forecast} & {\footnotesize{}MW, hourly} & {\footnotesize{}Wind infeed forecast for Germany published by transmission
system operators (TSO) at 8:00 d-1} & {\footnotesize{}European Energy Exchange (EEX),}{\footnotesize \par}

{\footnotesize{}\href{https://www.eex-transparency.com/}{https://www.eex-transparency.com/}} & {\footnotesize{} mlog, hourly value for all QHs}\tabularnewline
\hline 
\end{tabular}
\par\end{centering}
\caption{Overview of applied explanatory variables, their characteristics and
how to obtain them for the sake of reproducibility.}
\end{table*}
 Volume analysis has shown the importance of EPEX hourly auction prices.
Around noon, these prices mark the benchmark for any spot trading
activities. They provide an essential price indication for day-ahead
trading. Possible impacts on this market are expected to have a partial
influence on the intraday market as well.\\
\hspace*{0.5cm}All external determinants and their data sources are
summarized in Table 2. The calculations are made separately for every
quarter-hour of the day. Such a method shrinks the size of all matrices
in the calculation by 96 and reduces the computational effort immensely.
On the other hand, quarter-hourly interdependencies evoked by ramping
costs or similar load events are lost. Traditional thermal power plants
have boundaries like start-up times. These might cause one quarter-hour
to be profoundly affected by the preceding one. A principal component
analysis (PCA) acknowledges these effects in 
\begin{equation}
y_{h,t-1}\backsim\mathbf{\mathbf{\Lambda_{\mathrm{l,t}}}F_{\mathrm{l,t}}},
\end{equation}
where $\mathbf{\Lambda_{l,t}}$ are the load factors and $\mathbf{F_{l,t}}$
the principal components of all 96 prices of today's EXAA results,
today's EPEX day-ahead result and yesterday's lagged prices of the
market to be predicted. The components shall comprise all daily price
information and are determined using all 96 quarter-hours. Please
note that $l=1,...,96$ because 96 quarter-hours yield 96 components.
We run the PCA over the EXAA and EPEX day-ahead prices since they
are already available around 10:21 and 12:42 the day ahead and might
give a good indication of the most current price interdependencies.
In case of EPEX intraday continuous forecasts, we add a PCA on EPEX
QH prices based on the same argument and data availability. In addition,
a forth PCA on lagged prices tries to capture intraday dependencies
in the markets we aim to predict. As with conventional PCA, the first
few factors comprise sufficient information to be included. In our
case, three components are utilized. \\
\hspace*{0.5cm}While the ENTSO-E load forecast itself is already
expected to contain a good portion of price information, its connected
historical time series could deliver additional hints. Suppose that
a specific load profile determines the shape of quarter-hourly demand.
If we can identify days with a similar load curve, their observable
prices provide valuable input for our forecasts. This idea was used
in a comparable pre-filtering set-up by \citet{maciejowska2016probabilistic},
one of the winning teams in a price forecasting challenge. We will
likewise exploit this thought and aim to locate a similar load day\footnote{For a correct parameter identification, the actual process is twofold.
First, the calculus is carried out for historical data to retrieve
past same day prices for model tuning. In a second phase, the determination
is done for d+1 to have a valid input parameter for a live forecast
of prices. } from which to extract prices. The identified price will serve as
another input feature. We aim to extract a vector out of our feature
matrix that best approximates the day to be predicted with regards
to its Euclidean distance. In other words, the Euclidean distance
between the current day and all historical load observations is measured,
and the minimum is determined. Once found, the prices of the most
similar load scenario are plugged into the model assuming that they
inherit crucial information about upcoming price developments. \\
\hspace*{0.5cm}Regarding timing, we do not use any updated forecast
data, i.e., intraday predictions are made at the same point in time
that the QH auction prices are being estimated even though their computation
is not restricted to fixed auction times. This is essential because
we want to derive a coinstantaneous trading decision from the predictions,
i.e., enter positions in both markets at the same time. However, it
leads to a situation in which we use the most current data only for
the QH auction. It is a trade-off for the sake of publicly available
data and simultaneous applications of both forecasts to capture economic
benefits.

\subsection{Prediction model}

The aim is to predict both the EPEX quarter-hourly intraday auction
and the intraday continuous market price of the next day. An equivalent
model is utilized for both markets which is why the following notations
have a general character and are not restricted to one of the exchanges.
Our deliberations start with a plain benchmark model, denoted as \textbf{Naive}\textsubscript{EXAA}
in the rest of this paper. Whilst in other market regimes the best
naive guess is provided by yesterday's price, the German market offers
an idiosyncrasy in the form of the EXAA auction and its first indication
for later auctions and continuous trading to follow. We exploit the
EXAA results and expect them to be the best estimator for the other
markets such that $\hat{y}_{qh,t}=y_{\mathrm{EXAA},qh,t}$. This model
shall serve as an accuracy baseline for the other forecast approaches.
\\
\hspace*{0.5cm}Linear concepts tend to show convincing results in
energy forecasting (see \citet{maciejowska2016hybrid} for an example),
which is why this paper sets the technical focus on them. Of course
we could have used other predictors, like non-linear ones, but have
decided to thoroughly introduce the overall model architecture instead
of applying a wider set of models. For more information on other common
forecasting approaches and their accuracy one might refer to \citet{gurtler2018forecasting}.
With reference to the described input factors, we introduce two general
regression approaches that serve as a basis for all upcoming models.
Our first input set, denoted by the prefix \textbf{Expert}\_, takes
expert decisions on weekly dummies and lags and is described in the
following simplified form exemplarily for $y_{qh,t}$= EPEX quarter-hourly
auction quotation
\begin{align}
y_{qh,t} & =\beta_{qh,0}+\sum_{\substack{j\in\{1,2,7\}\\
i=(1,...,3)
}
}\underbrace{\beta_{qh,i}y_{qh,t-j}}_{\mathrm{\textrm{AR-terms}}}+\underbrace{\beta_{qh,4}\mathrm{\phi_{\mathit{1,qh,t}}}}_{\mathrm{\textrm{EEX wind}}}\\
 & +\underbrace{\beta_{qh,5}\mathrm{\phi_{\mathit{2,qh-1,t}}}}_{\mathrm{\textrm{EEX wind}lag}}+\mathbbm{1}_{(29,...,76)}(qh)\underbrace{(\beta_{qh,6}\mathrm{\phi_{\mathit{3,qh,t}}}}_{\mathrm{\textrm{EEX PV}}}+\underbrace{\beta_{qh,7}\mathrm{\phi_{\mathit{4,qh-1,t}}})}_{\mathrm{\textrm{EEX PV lag}}}\nonumber \\
 & +\sum_{\substack{k\in\{0,1,2,7\}\\
i=(1,...,4)
}
}\underbrace{\beta_{qh,7+i}y_{\mathrm{DA},qh,t-k}}_{\mathrm{\textrm{EPEX DA lags}}}+\sum_{\substack{k\in\{0,1,2,7\}\\
i=(1,...,4)
}
}\underbrace{\beta_{qh,11+i}y_{\mathrm{EXAA},qh,t-k}}_{\mathrm{\textrm{EXAA QH lags}}}\nonumber \\
 & +\underbrace{\beta_{qh,16}y_{\min,t-1}+\beta_{qh,17}y_{\textrm{max},t-1}}_{\mathrm{\textrm{non-linear\,effects}}}+\underbrace{\beta_{qh,18}\mathrm{\phi_{\mathit{5,qh,t}}}}_{\mathrm{\textrm{ENTSO load}}}+\underbrace{\beta_{qh,19}\mathrm{\phi_{\mathit{6,qh-1,t}}}}_{\mathrm{\textrm{ENTSO load lag}}}\nonumber \\
 & +\sum_{\substack{l=(1,...,3)\\
i=(1,...,3)
}
}\underbrace{\beta_{qh,19+i}\textrm{PC}\textrm{A}_{\mathrm{EXAA,}l}}_{\mathrm{\textrm{daily PCA factors}}}+\sum_{\substack{l=(1,...,3)\\
i=(1,...,3)
}
}\underbrace{\beta_{qh,22+i}\textrm{PC}\textrm{A}_{\mathrm{EPEX\,DA,}l}}_{\mathrm{\textrm{daily PCA factors}}}\nonumber \\
 & +\sum_{\substack{l=(1,...,3)\\
i=(1,...,3)
}
}\underbrace{\beta_{qh,25+i}\textrm{PC}\textrm{A}_{\mathrm{EPEX\,QH,}l}}_{\mathrm{\textrm{daily PCA factors}}}+\sum_{\substack{m=\{1,6,7\}\\
i=(1,...,3)
}
}\underbrace{\beta_{qh,29+i}D_{m}}_{\mathrm{\textrm{daily dummies}}}\nonumber \\
 & +\underbrace{\beta_{qh,30}\mathrm{y_{similiar\mathit{,qh,t}}}}_{\mathrm{\textrm{similar load day}}}+\varepsilon_{qh,t},\nonumber 
\end{align}
with $y_{qh,t-k}$ being the mlog prices of the identical quarter-hour
one, two and seven days ago and $y_{EXAA,qh}$, $y_{DA,qh}$ its equivalent
lags for the EXAA and EPEX day-ahead market. Obviously, the AR-term
changes with the market to be predicted. The terms $y_{\min,t-1}$
and $y_{\textrm{max},t-1}$ refer to yesterday's minimum and maximum
mlog price and are supposed to reflect the non-linear interdependency
between the daily price regimes, while $\phi_{\mathit{1,qh,t}},...,\phi_{\mathit{6,qh,t}}$
are the wind, PV and load forecasts for the respective delivery day
and its lagged values. We use the previous hours' lagged values to
capture ramp-up effects of our fundamental variables. The notation
$y_{\mathrm{similar},qh,t}$ describes prices of the minimum Euclidean
distance load scenario as mentioned in the previous sub-chapter, i.e.,
prices of a day that feature a similar load profile with regards to
the Euclidean distance between the current load forecast and all historical
ones. \\
\hspace*{0.5cm}The term $D_{k}$ is a dummy variable (i.e., taking
a value of 1 in case of occurrence) to capture the intra-week term
structure with $m=1,6,7$ for Monday, Saturday and Sunday. Weekly
seasonality is a crucial factor for spot electricity prices like the
ones present (see also \citet{weron2008forecasting} for an example
on three weekly dummies). Saturday and Sunday differ from the rest
of the week due to their weekend characteristics, with less traders
being active and lower load and energy production levels. Our markets
might be traded day-ahead, so even Monday could differ from typical
weekdays due to the fact that quantities were traded on a Sunday.
The argument certainly holds true for the day-ahead traded QH auction
and intraday continuous markets are at least partially traded one
day in advance. We therefore apply the set-up on both markets. The
notation $\textrm{PC}\textrm{A}_{\mathrm{EXAA},l}$ defines the $l-\mathrm{th}$
principal component of the EXAA QH prices. Besides EXAA, we include
PCA's for EPEX QH and EPEX day-ahead prices. The error term $\varepsilon_{h,t}$
is assumed to be independent and identically distributed (iid) with
$\varepsilon_{h,t}\sim N(0,\sigma_{qh}^{2})$. In case of EPEX intraday
continuous prices we slightly expand Eq. (4) by adding its relevant
auto-regressive lags, the current EPEX QH auction price as well as
a PCA on the intraday continuous prices. They are available before
the continuous trading window starts so it makes sense to exploit
them for forecasting models. Please note that our model in Eq. (4)
is a multivariate one meaning that we have an independent estimation
per quarter-hour or, in other words, 96 autarkic models.\\
Using expert decisions inevitably means subjectivity and leaves room
for criticism. We include a second input set called \textbf{Full}\_
that overcomes all concerns over possible bias. Instead of weekdays
for Monday, Saturday and Sunday the full model implements dummies
for every day of the week (such that $m=1,...,7)$ in equation (4).
It also includes all 7 lags for every quarter-hourly price compared
to the expert model only using lag 1,2 and 7. Lastly the full model
replaces all PCA's with 96 prices per quarter-hour for EXAA, EPEX
QH, EPEX day-ahead and -in case of the intraday continuous prices
to be predicted- for EPEX intraday. This expansion causes the model
structure to be much more complex than before. The full model features
254 predictors in case of QH auction predictions and over 300 for
intraday continuous forecasts. Such an expansive model might serve
as a sensitivity check. If our expert decisions are correct, than
the models shall result in similar accuracy. \\
\hspace*{0.5cm}The $\beta_{qh,1,...,30}$ parameters in Eq. (4) are
determined by the well-known ordinary least squares (OLS) optimization
in our first model, leading to the estimator called \textbf{LM} hereinafter.
One of the key points of this paper is an evaluation of the ideas
in \citet{ziel2015forecasting} and \citet{ziel2017modeling}. Does
the EXAA price add accuracy gains in QH markets? We introduce a second
model, \textbf{LM}\textsubscript{EXAA,} with one slight difference
to Eq. (4). All parameters remain unchanged for the prediction of
both intraday and auction markets, but we add the EXAA quarter-hourly
auction results as another explanatory variable. The sources above
found evidence for accuracy gains once EXAA prices were included,
which is why we expect them to enhance our models in a similar fashion.
\\
\hspace*{0.5cm}Another concern indirectly arises from Eq. (4). We
use a large set of input factors where many features are assumed to
be correlated. We apply a PCA but include a selection of lagged values
which are again inputs for the PCA. Hence, high correlation in our
predictors needs to be taken into account together with the fact that
too many variables could cause overfitting. A second linear prediction
model, denoted as \textbf{EN}, shall overcome this limitation. Introduced
in \citet{zou2005regularization}, elastic nets (EN) balance between
linear and quadratic penalty factors or between lasso and ridge regression.
Its great advantage is that it combines aspects out of the latter
two algorithms, such that elastic nets can automatically remove unneeded
variables entirely from the model while also being more robust to
correlation than the lasso. We simplify the model in Eq. (4) to the
regression form
\begin{equation}
y_{qh,t}=\sum_{j=1}^{p}\beta_{qh,t,j}x_{qh,t,j}+\varepsilon_{qh,t}.
\end{equation}
The OLS optimization aims to minimize the residual sum of squares
(RSS). The elastic net estimator expands this approach by adding a
linear penalty factor $\lambda_{qh}\geq0$ in
\begin{equation}
\hat{\mathbf{\mathbf{\mathrm{\boldsymbol{\beta}}}}}_{\mathbf{EN}}=\argmin_{\mathbf{\boldsymbol{\beta}}_{\mathit{\mathbf{qh}}}}\left\{ RSS+\underbrace{\lambda_{qh}\left(\frac{1-\alpha}{2}\sum_{j=1}^{p}\beta_{qh,j}^{2}+\alpha\sum_{j=1}^{p}\left|\beta_{qh,j}\right|\right)}_{\mathrm{\textrm{Penalty Term}}}\right\} ,
\end{equation}
\begin{align*}
\mathrm{where\;RSS}=\sum_{t=1}^{T}(y_{qh,t}-\sum_{j=1}^{p}\beta_{qh,j}x_{qh,t,j})^{2}.
\end{align*}
In case of $\lambda_{qh}=0$ we obtain the same results as for the
OLS-based LM model. The other extreme case $\lambda_{qh}\rightarrow\infty$
causes all variables to be shrunken to zero, i.e., removed from the
model or tending to zero depending on the weighting between lasso
and ridge regression. The allocation between ridge and lasso is described
by the parameter $\alpha\text{\ensuremath{\in}}\text{[0,1]}.$ We
follow the findings of an empirical study in \citet{uniejewski2016automated}
and set $\alpha=0.5$ as subjective expert decision, that is justified
by good predictive performance reported in the literature. The optimization
itself might be seen as a trade-off between minimizing the RSS and
simplifying the model structure. Besides, an elastic net is a form
of variable selection due to its ability to cancel out entire input
factors. A regularization method such as the elastic net urgently
necessitates normalization and standardization to yield valid results.
The penalty term works by both scale and magnitude of the variables
while we desire a sparse solution based solely on the individual magnitude.
However, the topic of standardization is of no concern in our context
since the mlog transformation explicitly regards this aspect. Please
note that in case of standardization there is no necessity for an
intercept anymore which is why there is none in Eq. (4).\\
\hspace*{0.5cm}Equation (6) leaves an optimization problem to be
solved. We compute a solution using R's \texttt{glmnet} package by
\citet{friedman2010regularization}. The optimization computation
requires a measure to be minimized, and in our case that is the mean
squared error (MSE). Based on a user-specified number of 1,000 different
steps for $\lambda_{qh}$, \texttt{glmnet} automatically creates
an exponential grid starting from $\lambda=0.001$ to a data-derived
maximum per each quarter-hour and determines the best value based
on a 10-fold cross-validation. Despite being more time intensive than
a simple optimization, our cross-validation set-up provides generalization
with regards to the selected $\lambda_{qh}$. Just like the previous
OLS model, a second predictor \textbf{EN}\textsubscript{EXAA} comprises
the quarter-hourly EXAA quotations of the delivery date.

\section{Back-test results\label{sec:4}}

\subsection{Point forecast performance}

\begin{figure*}
\centering{}\includegraphics[scale=0.6]{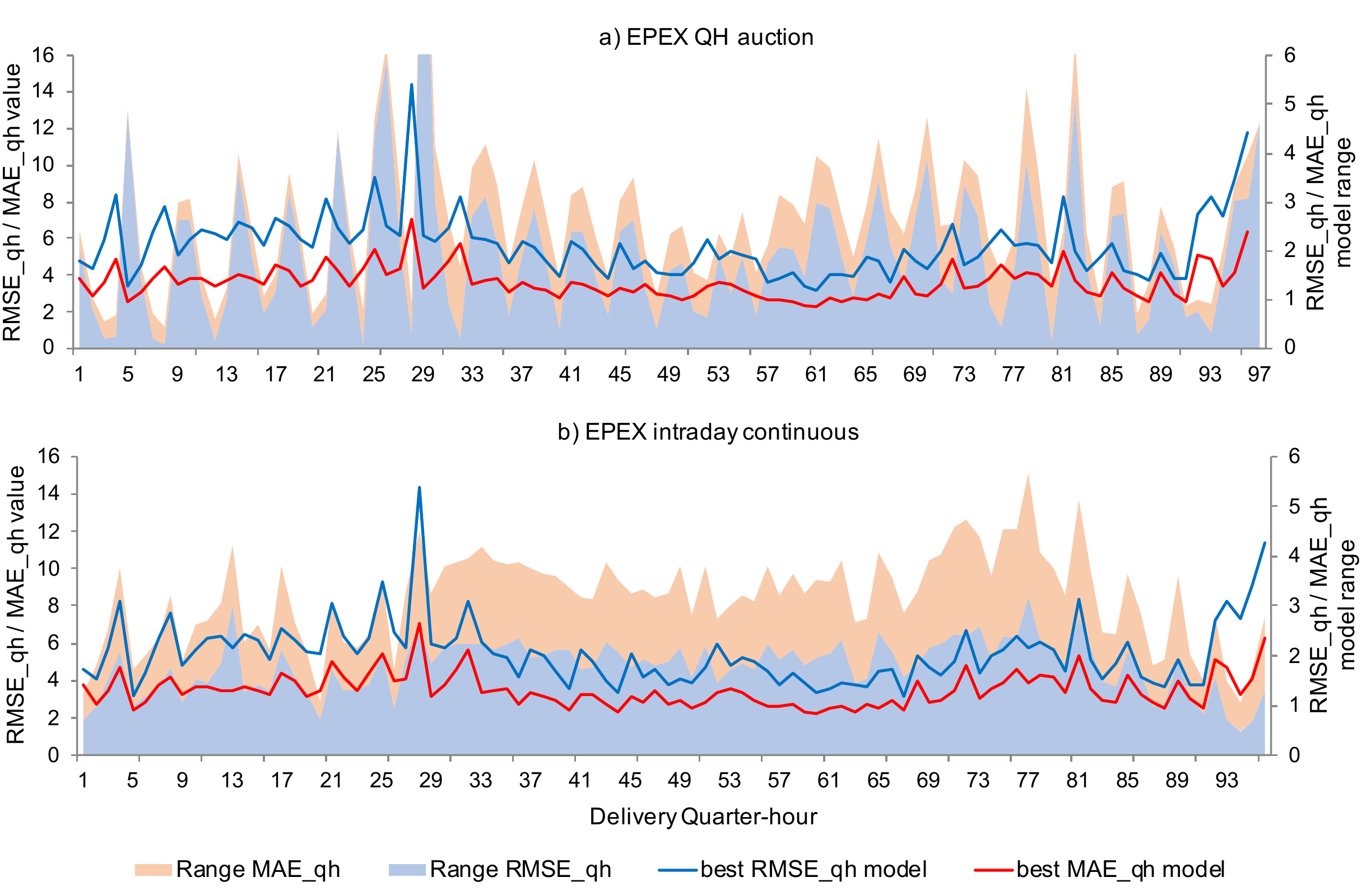}\caption{Quarter-hourly model fit metrics MAE\_qh and RMSE\_qh and range of
MAE\_qh and RMSE\_qh between best and worst model. The plot is limited
to the best performing model per market, in case of the QH auction
that is Full\_EN and for the intraday continuous market it is Expert\_EN.
Please note that we have excluded Full\_LM and Full\_LM\protect\textsubscript{EXAA}
from the plot due to its unreasonably high error metrics.}
\end{figure*}
Before turning the attention to economic gains stemming from accurate
forecasts, the predictive performance of our models in question requires
discussion. Rolling estimations assure realistic simulation results.
Hence, every model is iteratively fitted and predicts on new data,
while afterward the entire data matrix is shifted by 96 quarter-hours.
This modus operandi ensures that all predictions are made on out-of-sample
data and reflects realistic behavior in practical applications. We
train our model with nearly one year of data so that a period spanning
from 08.10.2015 to 06.10.2016 is utilized for the initial training.
From 07.10.2016 to 31.05.2017 all values are predicted in an out-of-sample
manner such that we have 57,714 individually estimated quarter-hours
to be evaluated in all upcoming tests. Given this vast amount of data,
we believe the test results to be sound.\\
\hspace*{0.5cm}We report two commonly used measures, the root-mean-square-error
(RMSE) and mean-absolute-error (MAE), given by 
\begin{equation}
\mathrm{RMSE}=\sqrt{\frac{1}{96T}\sum_{t=1}^{T}\sum_{qh=1}^{96}(y_{qh,t}-\hat{y}_{qh,t})^{2},}
\end{equation}
\begin{equation}
\mathrm{MAE}=\frac{1}{96T}\sum_{t=1}^{T}\sum_{qh=1}^{96}(\left|y_{qh,t}-\hat{y}_{qh,t}\right|),
\end{equation}
where $T$ describes the number of days, $y_{qh,t}$ the observed
prices and $\hat{y}_{qh,t}$ its predicted counterpart. All results
are reported in Table 3.
\begin{table}
\centering{}\includegraphics[scale=0.65]{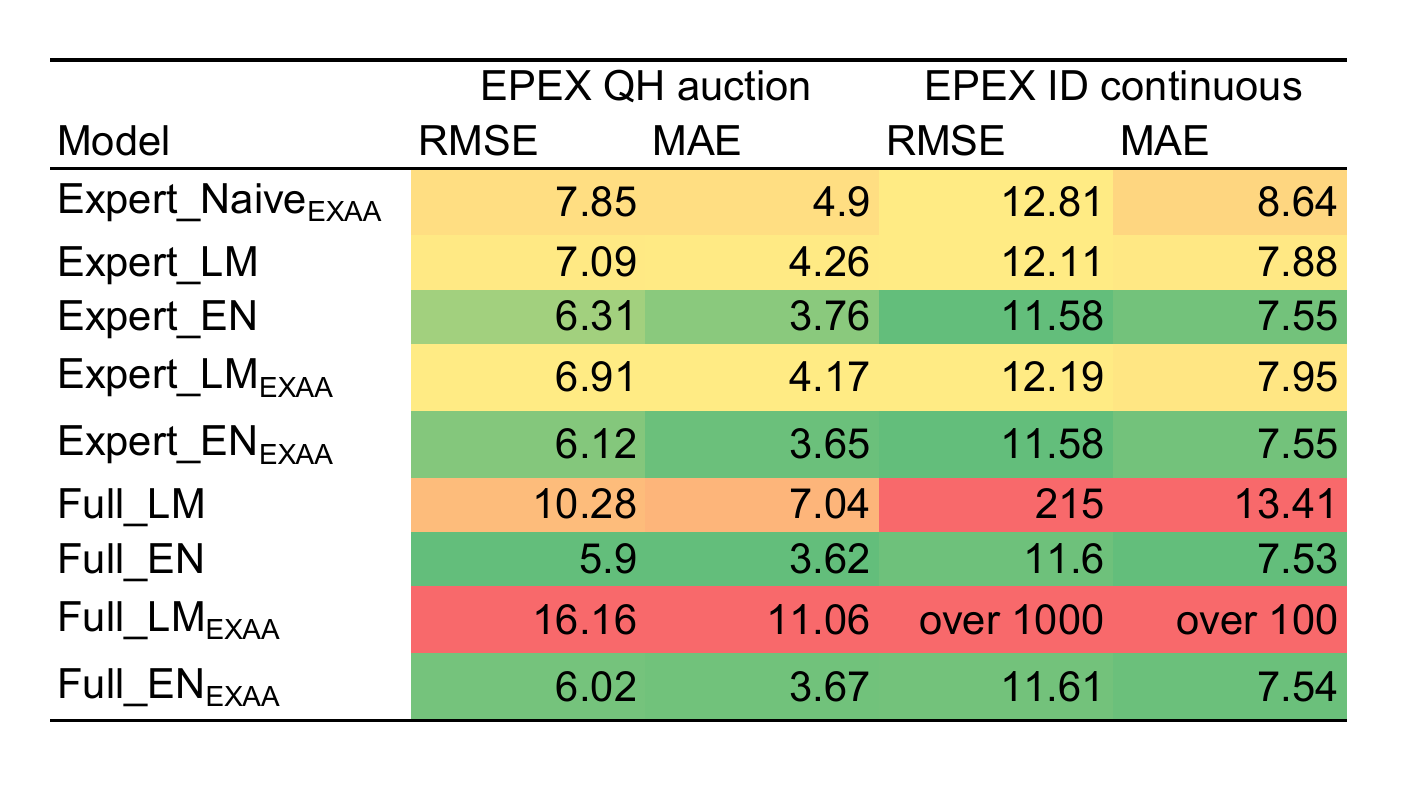}\caption{Error measures root-mean-square-error (RMSE) and mean-absolute-error
(MAE) for applied forecast models. }
\end{table}
 They suggest that the quarter-hourly auction indeed benefits from
forecasts based on external factors since the difference between the
benchmark model and the best performing EN estimator is more than
20\% in the RMSE case. The LM model is better than the naive benchmark,
and the elastic net approach tops that by roughly the same accuracy
gain that separated the LM and the naive model for both RMSE and MAE.
Given our range of auction data, advanced linear modeling seems to
add a crucial portion of performance. Interestingly, our choice of
expert decision was not entirely correct since the full models feature
lower MAE and RMSE values. However, this is not the case with LM models.
As expected, they cannot handle the massive set of inputs and feature
the highest RMSE and MAE results when enriched with all inputs.\\
\hspace*{0.5cm}At the same time, the EXAA as a model input leaves
the impression of minor importance. EXAA-enriched EN models outperform
their rivals by around 3\% for the QH auction if we consider the RMSE.
Still, this effect has been expected to be higher and is only limited
to the elastic net that can handle numerous input factors. The common
OLS-based LM model rather seems to suffer from more inputs. The EXAA
provides a quarter-hourly quotation for the same delivery date but
only slightly improves the models. This could either be caused by
the time lag from result publication at 10am to EPEX bidding at 3pm
or the different intraday characteristics respectively. Indeed, one
might argue that 5 or more hours could lead to new wind forecasts
and changed QH bids. Another thought is connected to the hourly day-ahead
auction. Presumably, market participants wait for the most important
German spot auction until they actively trade-out their quarter-hourly
shapes. Thus, the EXAA auction could be characterized by different
market players and changing bidding behavior. However, these thoughts
require quantitative backing in further research. \\
\hspace*{0.5cm}The picture changes with the EPEX continuous intraday
market. The performance is almost two times worse than QH auction
results in case of EN predictions. Both MAE and RMSE are considerably
higher for intraday estimations. An initial guess might be that this
observation is associated with even more substantial intra-model deviations.
However, the results suggest a different outcome. The linear models
only slightly increase the performance in comparison with the usage
of plain EXAA prices as a prediction model input. Whereas elastic
net outperforms the OLS-based predictor in QH auctions, its performance
gain is only marginal in the intraday regime. Our empirical test suggests
the same results for the question of EXAA influences on performance.
An accuracy increase of around 1\% does not support the argument of
strong EXAA implications in continuous intraday markets. The market
itself features an entirely different pricing regime which tends to
either be more complicated in prediction or influenced by other parameters.
Besides that, the timing aspect also matters. All intraday forecasts
exploit the same fundamental data that was used for the QH auction.
Updated wind or load data could boost accuracy. In terms of input
factors our expert models are comparable to the full models with one
exception; while the linear model was already struggling for QH auction
data, the intraday continuous trading proves it to be unsuitable for
large numbers of regressors. Its error measures clearly indicate a
model issue and unreasonable point forecasts. \\
\hspace*{0.5cm}Figure 4 provides a graphical representation of the
model fit. Please note that we change eq. (7) and (8) to a quarter-hourly
representation by adding the identically named suffix. It shows the
quarter-hourly term structure of the best performing MAE\_qh and RMSE\_qh
model as well as the range between the best and worst performing model.
It appears that each hour's last quarter-hour is harder to estimate
with higher RMSE\_qh and MAE\_qh results. This results in a characteristic
zig-zag pattern in both markets. Besides, the transition phase from
off-peak to peak between hour 7 and 8 and hour 20 to 21 is a common
time of higher uncertainty. Additional plants are ramped up to cover
tradeable peak profile demands. These effects are observable in higher
error measures in Figure 4. The overall QH auction's error range is
constant besides the last QH and off-peak/peak changes but the intraday
continuous plot reveals higher model deviations for the entire peak
time. So this market appears to be more difficult to predict in peak
hours.\\
\hspace*{0.5cm}
\begin{figure*}
\begin{centering}
\includegraphics[scale=0.7]{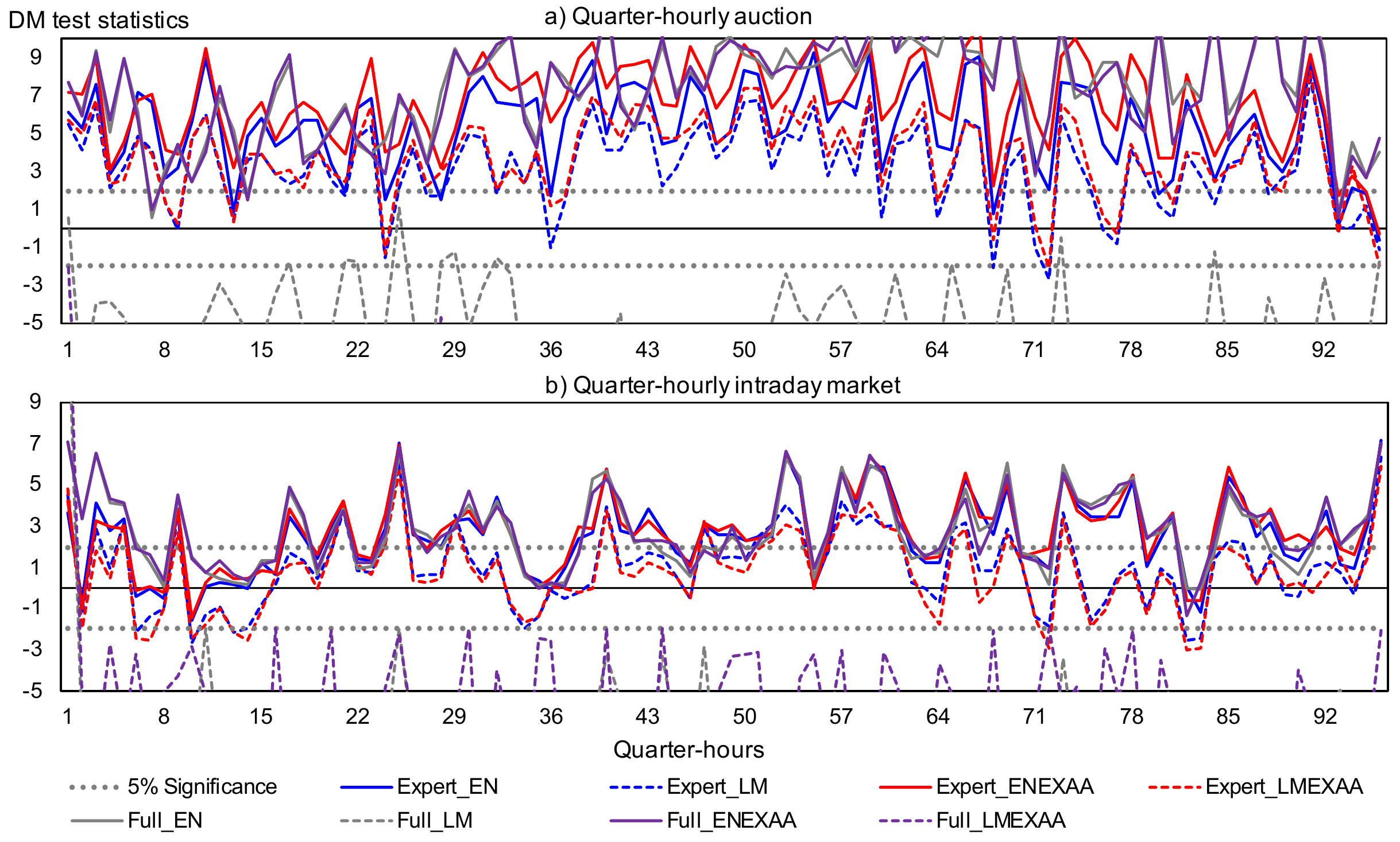}
\par\end{centering}
\caption{Quarter-hourly Diebold-Mariano test statistics carried out under the
absolute loss function and with loss series lagged four times determined
by an AR(p) process. }
\end{figure*}
A more advanced test measure is delivered by \citet{diebold2002comparing}
in the eponymous Diebold-Mariano (DM) test statistics. It has proven
to be a profound measure with energy pricing applications in \citet{nowotarski2014empirical}
and \citet{bordignon2013combining} and aims at investigating the
outperformance of one model forecast over the other. The test input
parameters are given by the loss differential series $\Omega_{qh,t}^{m1,m2}$
of the absolute error of models $m_{1},m_{2}$ such that
\begin{equation}
\Omega_{qh,t}^{m1,m2}=\left|y_{qh,t}^{m1}-\hat{y}_{qh,t}^{m1}\right|^{p}-\left|y_{qh,t}^{m2}-\hat{y}_{qh,t}^{m2}\right|^{p}.
\end{equation}
Depending on the choice of $p$, the quadratic loss or the absolute
loss is applied. Our tests did not reveal any considerable difference
in the test results for either $p=1$ or $p=2$ which is why we stick
to the former. An essential prerequisite of the test is non-covariance
stationarity in errors as discussed in \citet{diebold2015comparing}.
Daily test statistics might contradict this postulation since all
of the quarter-hours are driven by the same daily fundamental drivers
as proposed by \citet{nowotarski2016improving}. Our univariate approach
eludes this matter by its finer resolution. Another concern arises
from autoregressive structures. Since we include at least three lags,
the quarter-hours and their connected prices must be correlated. This
issue is dealt with by using lagged errors for Eq. (9). We inspect
the partial auto-correlation function and fit an AR(p) process to
the intraday and QH auction time series (see \citet{ziel2015forecasting}
for the idea of fitting an AR(p) process to tackle correlation in
the DM test) to identify the most suitable lag order. In our case,
an error series lagged four times appears to be statistically sound.
The test itself is performed at the 5\% significance level and reflects
consistent outperformance against the naive benchmark model. \\
\hspace*{0.5cm}
\begin{figure*}
\centering{}\includegraphics[scale=0.7]{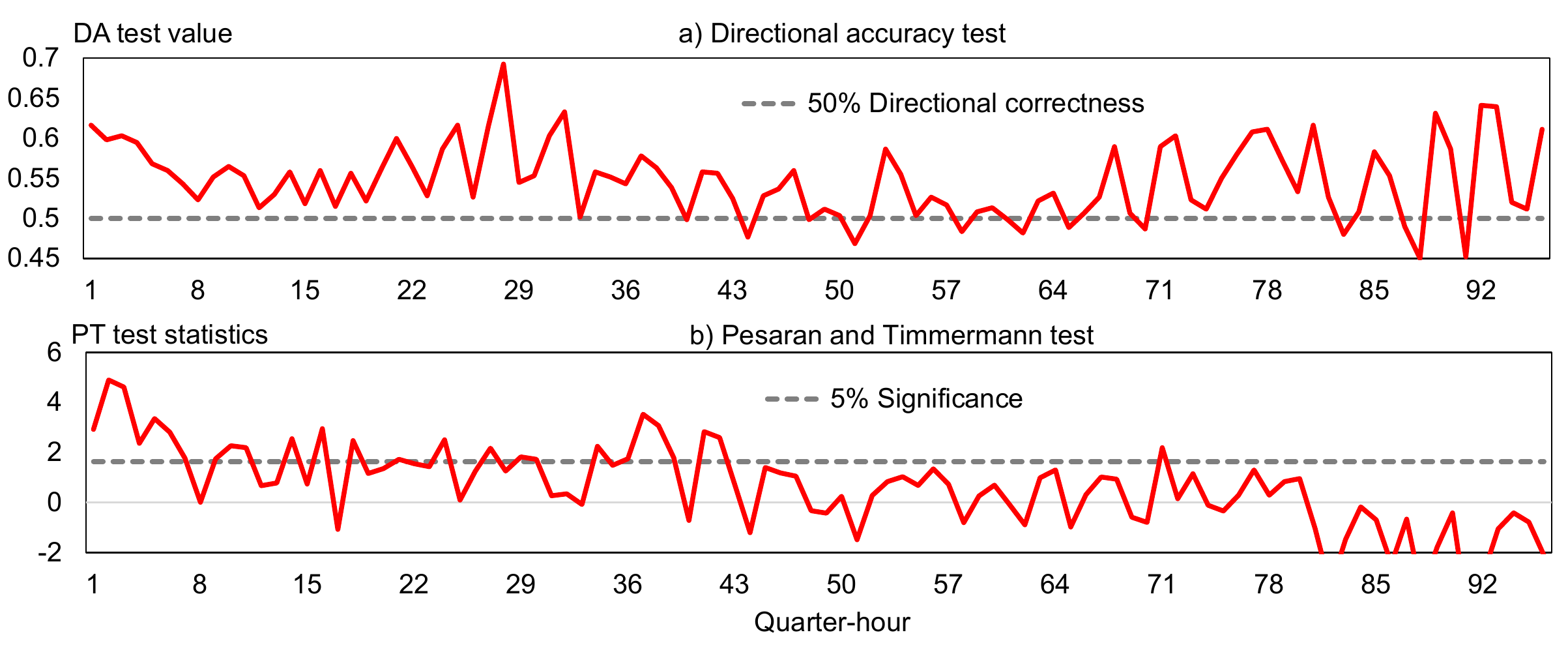}\caption{Directional forecast evaluations based on a) Directional Accuracy
statistics and b) the Pesaran and Timmermann test. The term 'directional'
describes whether the set of forecasts for the EPEX QH auction and
ID continuous market is capable of identifying the high price and
low price trading venue. }
\end{figure*}
 Figure 5 provides a graphical representation of the DM test results.
The higher the test statistics for every quarter-hour are, the better
the model performs in comparison with the benchmark model. Furthermore,
all values under or above the dotted gray line depict significant
overperfomance or underperformance of the respective model. Bearing
this in mind, Figure 5 supports the conclusion drawn from the RMSE
scores. Nearly all linear models with expert choices tend to improve
forecast accuracy for the QH auction with the EXAA-enriched ones better
than non-EXAA predictions, and EN estimates slightly more precise
than its OLS opponent. The LM models show significantly negative performance
compared to the benchmark, which again highlights their inability
to deal with larger amounts of regressors. All models seem to suffer
in the same period around QH 36. We can acknowledge that EXAA slightly
matters for the QH auction market based on our empirical study since
DM statistics are a bit higher for these models. Still, the effect
is very limited. The differences among the continuous intraday models
are reasonably low. Very few QHs show tendencies of statistical excess
performance, and even in these scenarios, it is difficult to favor
a specific model. The majority of observations are to be found in
the range below 5\% significance meaning neither LM, EN or EXAA enrichment
leads to fewer errors compared to our benchmark. This outcome was
unanticipated but might again be due to the time lag between estimations
and continuous intraday trading activities.

\subsection{Economic effects of accurate forecasts}

\subsubsection{Directional forecast portfolio approach}

A single point forecast has limited value if considered separately
without a translation into a trading decision. We will introduce two
different approaches that shall use the forecasts as an input and
transform these into a QH deal. Buying and selling are regarded in
different portfolios to reflect possible gains for net buyers and
sellers. Based on these thoughts, we firstly utilize both predictions
in a simplified binary scheme. Companies need to buy or sell their
residual quarter-hourly spot profile on spot exchanges and shall do
so based on the simple rule of buying in the cheaper market (low market)
and selling in the more expensive one (high market). Hence, a sell
position is entered into the market with higher predicted prices,
denoted as \textbf{Base}\textsubscript{Sell,} while the \textbf{Base}\textsubscript{Buy}
portfolio buys in the lower projected market. Since the previous sub-chapter
reflected an apparent tendency towards the EN predictors being the
best, we consider EN and EN\textsubscript{EXAA }for our analysis
and introduce additional portfolios, \textbf{Base}\textsubscript{Sell\_EXAA}and
\textbf{Base}\textsubscript{Buy\_EXAA}. These will explicitly include
the information provided by EXAA prices just as in the EN and LM forecast
models.\\
\hspace*{0.5cm}The above idea narrows the deal determination down
to a directional forecast based on the high and low market. Therefore,
we want to elaborate the directional accuracy of our approach. The
common measure (i.e., used in \citet{moosa2015directional}) Directional
Accuracy (DAcc) delivers the first hint of the binary accuracy of
our forecasts in a directional setting and is defined in a low market/high
market application as
\begin{equation}
DAcc=\frac{1}{n}\sum_{i=1}^{n}d_{qh,t},
\end{equation}
with the connected hit series 
\begin{equation}
d_{qh,t}=\begin{cases}
1, & \mathrm{if}\,(\hat{y}_{qh,t}^{m1}>\hat{y}_{qh,t}^{m2})\land(y_{qh,t}^{m1}>y_{qh,t}^{m2})\\
0, & \mathrm{if}\,(\hat{y}_{qh,t}^{m1}<\hat{y}_{qh,t}^{m2})\land(y_{qh,t}^{m1}>y_{qh,t}^{m2}).
\end{cases}
\end{equation}
Intuitively speaking, Eq. (11) assigns a value of 1 every time the
higher or lower market is correctly predicted. The representation
is kept general, but in our given case the model indices $m1,$ $m2$
denote either the EPEX QH auction or the QH intraday market. Once
we know whether the prediction of the higher market is right or wrong,
the DAcc measure in Eq. (10) reports the share of correct directional
estimates. The second framework is provided by \citet{pesaran1992simple}
and supposes independent directions of the observed and predicted
realizations under the null hypothesis, i.e., that estimated directions
do not add extra knowledge. Both metrics will be reported quarter-hourly
to gain additional insights into the time structure accuracy of the
predictions.\\
\hspace*{0.5cm}
\begin{figure*}
\centering{}\includegraphics[scale=0.7]{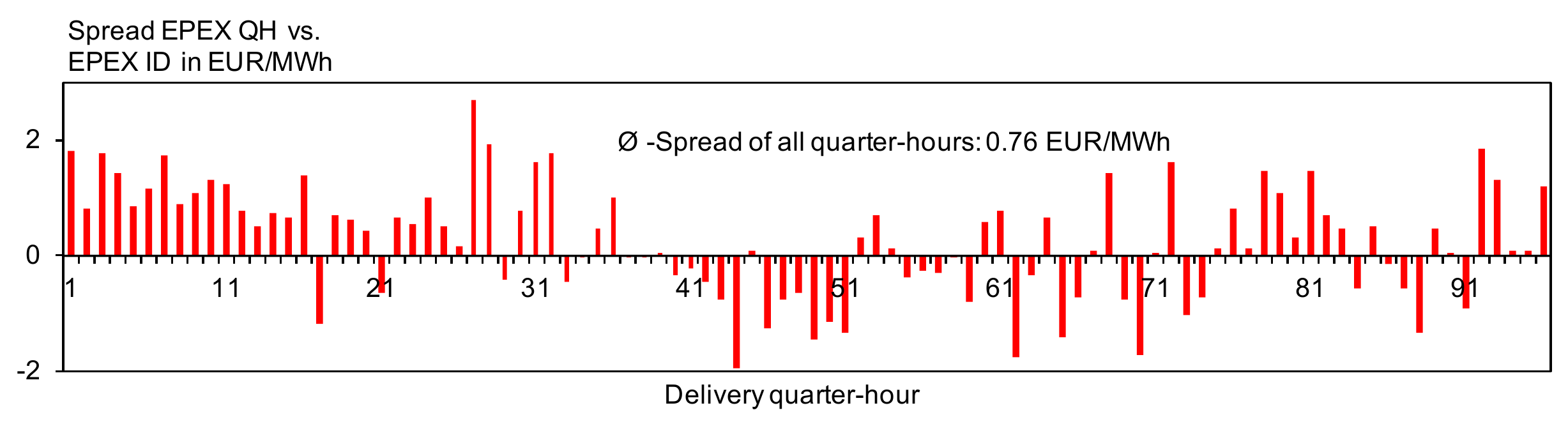}\caption{Quarter-hourly spreads of portfolio strategy Base\protect\textsubscript{Sell\_EXAA}/Base\protect\textsubscript{Buy\_EXAA},
i.e., sell in the predicted high market and buy in its lower equivalent.
The markets under consideration are the continuous QH intraday market
and the EPEX QH intraday call auction.}
\end{figure*}
\begin{table*}[!t]
\begin{centering}
\begin{tabular}[t]{>{\raggedright}m{0.3cm}>{\centering}p{2.4cm}>{\centering}p{6cm}>{\raggedright}p{1.2cm}>{\raggedright}p{1.2cm}>{\raggedright}p{1.2cm}>{\raggedright}p{1.2cm}>{\raggedright}p{1.2cm}}
 & {\footnotesize{}Portfolio ID} & {\footnotesize{}Description} & {\footnotesize{}Price} & {\footnotesize{}Min Price} & {\footnotesize{}Max Price} & {\footnotesize{}Std.Dev} & {\footnotesize{}Sharpe- Ratio}\tabularnewline
\midrule
\multirow{6}{0.3cm}{\footnotesize{}\multirow{1}{*}[1ex]{\rotatebox{90}{Benchmark  portfolios}}} & {\footnotesize{}Naive\textsubscript{EXAA}} & {\footnotesize{}EXAA QH trading of residual volumes} & {\footnotesize{}34.95} & {\footnotesize{}-102.34} & {\footnotesize{}168.85} & {\footnotesize{}17.18} & {\footnotesize{}2.04}\tabularnewline
 & {\footnotesize{}Naive\textsubscript{AUQH}} & {\footnotesize{}EPEX intraday QH auction trading of resuming position} & {\footnotesize{}34.66} & {\footnotesize{}-134.82} & {\footnotesize{}290.65} & {\footnotesize{}19.15} & {\footnotesize{}1.81}\tabularnewline
 & {\footnotesize{}Naive\textsubscript{IDQH}} & {\footnotesize{}EPEX intraday QH VWAP trading of residual position} & {\footnotesize{}34.81} & {\footnotesize{}-241.83} & {\footnotesize{}329.81} & {\footnotesize{}20.27} & {\footnotesize{}1.72}\tabularnewline
 & {\footnotesize{}Naive\textsubscript{reBAP}} & {\footnotesize{}Settlement of residual position at reBAP price with
grid operator} & {\footnotesize{}34.74} & {\footnotesize{}-2558.42} & {\footnotesize{}24455.05} & {\footnotesize{}147.54} & {\footnotesize{}0.24}\tabularnewline
 & {\footnotesize{}Perfect\textsubscript{Buy}} & {\footnotesize{}Full information benchmark portfolio, always buys
in lower market} & {\footnotesize{}30.74} & {\footnotesize{}-241.83} & {\footnotesize{}166.42} & {\footnotesize{}19.40} & {\footnotesize{}1.58}\tabularnewline
 & {\footnotesize{}Perfect\textsubscript{Sell}} & {\footnotesize{}Full information benchmark portfolio, always sells
in lower market} & {\footnotesize{}38.73} & {\footnotesize{}-117.77} & {\footnotesize{}329.81} & {\footnotesize{}19.22} & {\footnotesize{}2.02}\tabularnewline
\midrule
\multirow{8}{0.3cm}{\footnotesize{}\multirow{1}{*}[-1ex]{\rotatebox{90}{Forecast  portfolios}}} & {\footnotesize{}Base\textsubscript{Buy}} & {\footnotesize{}Buy in market with lowest predicted price using EN} & {\footnotesize{}34.38} & {\footnotesize{}-173.94} & {\footnotesize{}329.81} & {\footnotesize{}19.83} & {\footnotesize{}1.74}\tabularnewline
 & {\footnotesize{}Base\textsubscript{Sell}} & {\footnotesize{}Sell in market with highest predicted price using
EN} & {\footnotesize{}35.09} & {\footnotesize{}-241.83} & {\footnotesize{}266.17} & {\footnotesize{}19.61} & {\footnotesize{}1.78}\tabularnewline
 & {\footnotesize{}Base\textsubscript{Buy\_EXAA}} & {\footnotesize{}Buy in market with lowest predicted price using EN\textsubscript{{\footnotesize{}EXAA}}} & {\footnotesize{}34.36{*}} & {\footnotesize{}-173.94} & {\footnotesize{}329.81} & {\footnotesize{}19.84} & {\footnotesize{}1.73}\tabularnewline
 & {\footnotesize{}Base\textsubscript{Sell\_EXAA}} & {\footnotesize{}Sell in market with highest predicted price using
EN\textsubscript{{\footnotesize{}EXAA}}} & {\footnotesize{}35.11{*}{*}} & {\footnotesize{}-241.83} & {\footnotesize{}266.17} & {\footnotesize{}19.59} & {\footnotesize{}1.79}\tabularnewline
 & {\footnotesize{}MeanVar\textsubscript{Buy}} & {\footnotesize{}Mean-variance portfolio with lowest return, i.e.,
lowest price to pay using EN} & {\footnotesize{}34.71} & {\footnotesize{}-178.50} & {\footnotesize{}245.99} & {\footnotesize{}19.09} & {\footnotesize{}1.83}\tabularnewline
 & {\footnotesize{}MeanVar\textsubscript{Sell}} & {\footnotesize{}Mean-variance portfolio with highest return, i.e.,
highest price to sell using EN} & {\footnotesize{}34.76} & {\footnotesize{}-173.89} & {\footnotesize{}213.02} & {\footnotesize{}18.62} & {\footnotesize{}1.87}\tabularnewline
 & {\footnotesize{}MeanVar\textsubscript{Buy\_EXAA}} & {\footnotesize{}Mean-variance portfolio with lowest return, i.e.,
lowest price to pay using EN\textsubscript{{\footnotesize{}EXAA}}} & {\footnotesize{}34.72} & {\footnotesize{}-178.50} & {\footnotesize{}245.99} & {\footnotesize{}19.00} & {\footnotesize{}1.83}\tabularnewline
 & {\footnotesize{}MeanVar\textsubscript{Sell\_EXAA}} & {\footnotesize{}Mean-variance portfolio with highest return, i.e.,
highest price to sell using EN\textsubscript{{\footnotesize{}EXAA}}} & {\footnotesize{}34.76} & {\footnotesize{}-173.89} & {\footnotesize{}213.02} & {\footnotesize{}18.62} & {\footnotesize{}1.87}\tabularnewline
\bottomrule
\end{tabular}
\par\end{centering}
\caption{Empirical test results of different portfolio strategies in the case
study period from 07.10.2017 - 31.05.2018. The prices are not volume
weighted nor adjusted in any way and reflect the price one would buy
or sell at given the selected portfolio strategy. Naive prices denote
the simple average of the respective price series. Both the lowest
buy price ({*}) and the highest sell price ({*}{*}) are marked for
convenience.}
\end{table*}
Figure 6 summarizes the findings in a combined way. The upper plot
shows that using the individual forecasts to estimate the cheaper
or more expensive exchange leads to more than 50\% correctness in
most cases. In general, this is a promising finding since once we
have a higher correctness rate than 50\%, there is a possibility to
observe economic benefits. However, this postulation only holds true
if the losses of an incorrect prediction and the gains of a correct
one are equally distributed such that the cost of making a wrong prediction
is nearly equal to the benefit of being correct. On the other hand,
we see a decline in directional accuracy in the peak QHs ranging roughly
from quarter-hour 36 to 70. Our estimations seem to be more accurate
in off-peak regimes given the dataset. This message is supported by
the second metrics depicted in the lower part of Figure 6. The Pesaran
and Timmermann (PT) test statistics exhibit an off-peak/peak pattern.
The actual test score is contradictory to the measure mentioned before.
The majority of quarter-hours do not pass the test, meaning that we
found evidence that the correct direction and its predicted equivalent
are less independent than desired. This outcome was unforeseen considering
the results of the Directional Accuracy test. To conclude, the tests
suggest a promising rate of correctness but do not allow us to reject
the null hypothesis of autonomous directional errors. The forecast
quality might be biased. Still, we have to acknowledge that we only
want to investigate the economic value of our point forecasts and
have translated them into a binary framework. So, they could be distorted
since the basis is not a designated directional estimation.

\subsubsection{Mean-variance portfolio selection}

A different portfolio composition technique is given by mean-variance
portfolio selection. Initially introduced in \citet{markowitz1952portfolio},
its classical scope covers financial markets and the selection of
stocks under expected return and variance. However, there are a few
energy market applications of mean-variance concepts available (the
interested reader might refer to a recent review of this topic in
\citet{calvo2017energy}). To apply such, the definition of return
needs to be clarified. Financial markets assume a fixed asset position
with payments of price movements leading to a return given by $r_{\mathrm{traditional},qh,t}=(y_{qh,t}/y_{qh,t-1})-1$.
This notation makes sense for storable assets or long-term power contracts
but does not apply to a spot market example. Long-term contracts,
like futures, are usually settled daily in a margining process such
that only the price difference is paid or received. The same holds
true for a stock position. In spot markets, the daily position will
most likely be different due to changing off-take or power plant generation.
Hence, the resulting cash-flow is different. A consecutive two-day
long position of 50MW will not just be settled at the price delta
between day one and day two (as done with futures and daily margining),
but a market participant has to pay 50 MW times the market price.
Therefore, we will regard the price itself as the return leading to
our notation $r_{qh,t}=y_{qh,t}$. Another difference is given by
the differentiation into buy and sell portfolios. Once we value a
high return (or in our notation a high clearing price) as desirable
and optimize with regards to that, we will identify a sell portfolio
because a market player obviously demands high prices and high returns.
The buy portfolio is the inverse of the particular optimization result
and yields lower returns or lower prices for net buyers in the market.
\\
\hspace*{0.5cm}The mean-variance theory incorporates expected returns
and variance into an optimization framework. Individual assets numbered
by $i=1,...,n$ are weighted by a factor $w_{i,qh,t}$ to compose
a portfolio of assets. In our concrete case, the portfolio is restricted
to two assets or the choice between the QH intraday auction and continuous
intraday trading. Unlike financial applications, we do not include
any risk-free benchmark assets. Using our prices as single time series
returns in the Markowitz sense leads to a portfolio return in 
\begin{equation}
r_{\mathrm{portfolio,\mathit{qh,t}}}=\sum_{i=1}^{2}w_{i,qh,t}y_{i,qh,t},
\end{equation}
where $y_{i,qh,t}$ are the realized values for either the QH auction
or the intraday market and $w_{i,qh,t}$ are the connected weights.
Yet, Eq. (12) only provides insights into the historical return and
does not comprise any future-oriented quantification. Markowitz optimizations
require expected returns denoted as $E(r_{\mathrm{portfolio,\mathit{qh,t}}})$
which inevitably necessitates expected single $i$-th returns, i.e.,
$E(r_{i,qh,t})$$=E(y_{i,qh,t})=\mu_{i,qh,t}$. Instead of the traditional
mean formulation, we want to approximate the expected return by means
of our forecasts so that $\mu{}_{i,qh,t}\sim\hat{y}_{i,qh,t}.$ Pruning
the notation to just a single weighting factor and taking into consideration
the thoughts on expected return yields a more simple form such that
\begin{equation}
E(r_{\mathrm{portfolio,\mathit{qh,t}}})=\hat{y}_{1,qh,t}+w_{2,qh,t}(\hat{y}_{2,qh,t}-\hat{y}_{1,qh,t}).
\end{equation}
The variance is determined by a simplifying relaxation. Instead of
complex estimation schemes, we will apply the empirical\footnote{Please note that we apply the described rolling estimation shifts
to determine the empirical variance. Hence, the first window to calculate
the variance ranges from 08.10.2015 until 07.10.2016. This time span
is shifted by 96 units for every single day and ensures a unique empirical
variance for every day and every quarter-hour.} variance $\sigma_{i,qh,t}^{2}$ of the individual exchange return
series and assume it to be the best estimator in the calculation of
the portfolio return in 
\begin{equation}
\sigma_{\mathrm{portfolio},qh,t}^{2}=(w_{1,qh,t}^{2}\sigma_{1,qh,t}^{2})+(w_{2,qh,t}^{2}\sigma_{2,qh,t}^{2})+2w_{1,qh,t}w_{2,qh,t}\rho_{12,qh,t},
\end{equation}
with $\rho_{12,qh,t}$ being the correlation of the returns. We simplify
Eq. (14) to eliminate $w_{1,qh,t}$:
\begin{align}
\sigma_{\mathrm{portfolio},qh,t}^{2}=\sigma_{1,qh,t}^{2}+2w_{2,qh,t}(\rho_{12,qh,t}-\sigma_{1,qh,t}^{2})\\
+(w_{2,qh,t}^{2})(\sigma_{1,qh,t}^{2}-2\rho_{12,qh,t}+\sigma_{2,qh,t}^{2}).\nonumber 
\end{align}
An important part of portfolio theory is the identification of all
efficient portfolios under non-zero weights and a sum of weights equal
to one. The latter postulation results in the assumption of perfectly
divisible asset portions. We follow this traditional concept but must
acknowledge that under real-life trading circumstances the exchange
pre-defined minimum tick sizes condition small adjustments to the
optimization results due to the fact that they are not tradable. We
are also not interested in computing the entire set of efficient portfolios
but want to find the one portfolio that exhibits the highest utility
for the market participant. The utility function is defined as an
optimization problem in (basic form taken from \citet{calvo2017energy})
\begin{align}
U_{qh,t} & =\argmin_{\mathit{w_{2,qh,t}}}E(r_{\mathrm{portfolio,\mathit{qh,t}}})-\frac{1}{2}\gamma\sigma_{\mathrm{portfolio},qh,t}^{2}\\
\mathrm{s.t.\;}\; & w_{i,qh,t}\:\text{\ensuremath{\in}}\:[0,1],\nonumber 
\end{align}
in which $\gamma$ denotes a variable to specify the risk-aversion
of the market participant. We follow the energy literature and set
$\gamma=2$ which is regarded to be a slightly higher average risk
appetite (\citet{gokgoz2012financial,liu2007portfolio}). Given the
high variance of the intraday series an adjustment towards less risk
aversion appears to be suitable. Otherwise, the optimization will
mostly select the QH auction market. At the same time, the slight
changes to the original equations in Eq. (13) and Eq. (15) yield only
one weighting parameter $w_{2,qh,t}$ to be optimized. If we consider
that possible solutions are restricted to be anything between zero
or one, it becomes evident that we implicitly meet the requirement
$\sum_{i=1}^{2}w_{i,qh,t}=1$. We use R's standard optimization command
\texttt{optim} to find a solution for Eq. (16). The optimization
result yields two trading indications; if we value positive returns
as desired to sell at high prices, the portfolio \textbf{MeanVar}\textsubscript{Sell}
is the important one whereas its counterpart \textbf{MeanVar}\textsubscript{Buy}
sets the focus on negative returns and lower prices for a net buyer.
The same contentual separation counts for the EXAA-enriched equivalents
\textbf{MeanVar}\textsubscript{Sell\_EXAA} and \textbf{MeanVar}\textsubscript{Buy\_EXAA}
respectively.

\subsubsection{Economic portfolio assessment}

Now that we have determined different portfolio strategies with EXAA
and non-EXAA variations, the last facet to assess is the economic
gain or loss resulting from our underlying forecasts and portfolio
strategies. For the sake of simplicity, we neglect all kinds of fees
and trading charges as well as the price impacts possible bids might
have. Hence, we assume sufficient market liquidity to absorb additional
trading volumes. Last but not least, volume weighted average prices
(VWAP) are only an approximation for continuous market prices. Apparently,
a market participant does not have direct access to index quotations.
Instead, regular trading activities could lead to average deal prices
near the VWAP. Since the intraday trading activities are up to individual
counterparts, with a detailed time series not being available, we
apply the VWAP as a best guess. Based on these prices, we carry out
a simple portfolio simulation and check the average portfolio price
a market participant would pay or receive when following the portfolio
strategy. The back-test ranges from 07.10.2016 to 31.05.2018 and is
summarized in Table 4 together with a synopsis of all portfolio strategies.
We use the original prices to get the most realistic results. The
only adaptation we apply is the clock-change adjustment described
under sub-section 3.1. We acknowledge that this causes a small bias
but since it only accounts for two hours of each year we ignore the
clock-change in the trading simulation. Besides the usual standard
measures on time series resolution, we report a common portfolio management
criterion called Sharpe-Ratio (adjusted from \citet{calvo2017energy})
\begin{equation}
S=\frac{\frac{1}{96T}\sum_{t=1}^{T}\sum_{qh=1}^{96}(y_{\mathrm{strategy},qh,t})}{\sigma_{\mathrm{strategy}}},
\end{equation}
where the numerator describes the average realized price of the respective
portfolio strategy over all days and quarter-hours and $\sigma_{\mathrm{strategy}}$
the standard deviation of the realized prices of each strategy. The
strategy prices $y_{\mathrm{strategy},qh,t}$ are individually determined
per strategy, as previously described. In case of Base\textsubscript{Sell}
for instance, the strategy prices equal the market price of the higher
predicted exchange. Please bear in mind that in its conventional form
the Sharpe-Ratio applies the average excess return, but since we set
the risk-free rate to be equal to zero, this step is not necessary,
and the realized portfolio price is identical to the excess return.
\\
\hspace*{0.5cm}The naive portfolios only buy or sell in one market
at the simple average of the time series and consequently yield lower
sell and higher buy prices. There is no buy or sell separation with
the naive prices while the forecast approaches imply a buy and sell
market price. Consequently, our naive singular market strategies yield
no spread benefits. We likewise report a perfect portfolio strategy
under the assumption of complete market information. The results are
highly unlikely to be achieved in a real-world scenario but represent
the obtainable gains from fully accurate forecasts. However, we will
not discuss the perfect portfolio in depth but focus our attention
on achieved spreads compared to singular market activities as these
depict current market participant behavior more than the postulation
for complete ex-ante market knowledge. In general, the forecast portfolios
perform well. Our results point towards an outperformance of high/low
market interaction referred to as Base{\footnotesize{}\textsubscript{Buy}}
and Base{\footnotesize{}\textsubscript{Sell}} and their EXAA-enriched
equivalents. 

In detail, market participants buy 0.70 - 0.74€/MWh cheaper and sell
0.74 - 0.78€/MWh higher compared to any other of the individual markets.
Interestingly, the addition of EXAA prices yields higher spreads.
While the EXAA-aided point forecasts become only a bit more accurate
for the QH auction, the directional accuracy tends to improve. This
finding seems contradictory at first, but might be the case since
a directional forecast does not advance from a precise point prediction
but solely from correct high/low market estimates. \\
\hspace*{0.5cm}The Markowitz approach adds a considerably lower portion
of economic gains. Its portfolio structure is a trade-off between
the auction and continuous intraday prices. The realized portfolio
price varies between the QH auction and its continuous equivalent.
A possible explanation might be given by the Markowitz inputs. The
optimization has to split between the highly volatile intraday continuous
market and the more moderate QH auction. Most of the time, this results
in a significant portion of QH auction prices due to risk aversion
tendencies. Hence, if one considers the utility function in Eq. (16),
a more risk averse portfolio is created. While the plain prices do
not suggest larger benefits from following Markowitz-guided trading
in comparison with the base strategies, the Sharpe-Ratio and standard
deviation do. Both the Markowitz portfolio and the Sharpe-Ratio include
a variance measure in their calculus. Therefore, it does not come
as a surprise that the best Sharpe-Ratio results are provided by mean-variance
portfolios. Still, we would have expected at least a small portion
of economic benefit expressed in better spread levels. An explanation
for the performance is the concern over correlation. Our choice of
assets was predetermined, and we have not checked the correlation
between the time series, but in financial markets, the co-movement
among stocks contributes to a less balanced portfolio composition.
The picture might change with less correlation between assets. However,
the empirical results do not provide evidence for Markowitz approaches
to perform better regarding higher spreads but construct a risk-minimizing
portfolio. Therefore, we favor the simple base strategies that are
grounded on a high/low market scenario and will purely focus on such
in the detailed analysis.\\
\hspace*{0.5cm}A simple t-test depicted in Table 5 is supposed to
deliver further evidence on the statistical soundness of the identified
excess performance. The p-values propose significant differences between
our forecast-aided base portfolio prices and the intraday continuous
time series. The QH auction result is less clear and shows signs of
correlation with the non-EXAA base strategies. The result at least
partially confirms our findings. Forecast applications translated
into a simple buy/sell trading decision result in different portfolio
price means compared to the underlying individual prices. There are
tests available for the equality of Sharpe-Ratios. They use the portfolio
prices as inputs and check for statistically sound differences among
Sharpe-Ratios. We apply the classical pairwise test of \citet{ledoit2008robust}
and an expansion that considers joint effects of prices in a multiple
Sharp-Ratio test in \citet{leung2006testing} and later for non-iid
cases in \citet{wright2014test}. Results are reported in Table 5.
While the multiple test statistics clearly point towards independent
Sharpe-Ratios, some of the pairwise test findings have to be rejected.
However, this does not contradict our general statement of independent,
considerable differences in prices when using forecasts since most
of the combinations that appear to be correlated are using a slightly
changed set of inputs and might indeed be nearly equal. 
\begin{table}
\begin{tabular}{>{\centering}p{1.7cm}cccc}
\hline 
 & \multicolumn{4}{c}{{\footnotesize{}p-values}}\tabularnewline
 & {\footnotesize{}Base\textsubscript{Buy}} & {\footnotesize{}Base\textsubscript{Sell}} & {\footnotesize{}Base\textsubscript{Buy\_EXAA}} & {\footnotesize{}Base\textsubscript{Sell\_EXAA}}\tabularnewline
\hline 
{\footnotesize{}Naive\textsubscript{AUQH}} & {\footnotesize{}0.016} & {\footnotesize{}0.017} & {\footnotesize{}0.005} & {\footnotesize{}<0.001}\tabularnewline
{\footnotesize{}Naive\textsubscript{IDQH}} & {\footnotesize{}<0.001} & {\footnotesize{}<0.001} & {\footnotesize{}<0.001} & {\footnotesize{}0.005}\tabularnewline
\hline 
\end{tabular}\caption{T-test for statistical significance of lower buy and higher sell prices.
The two-sided test postulates $H_{0}:\mu_{1}-\mu_{2}=0$ and checks
for statistically sound differences in portfolio prices.}
\end{table}
\\
\hspace*{0.5cm}Table 4 implies homogeneity across all QHs. We additionally
want to analyze time structure effects on the economic outcome and
turn our attention to the realized spread of the best performing Base\textsubscript{Sell}/Base\textsubscript{Buy}
strategy. Based on the forecasts, we observe a high/low spread (the
delta between high and low prices) of 0.76€/MWh among all QHs. Figure
7 cascades this singular number into a finer granularity. It depicts
limitations for the peak-load ranging from QHs 32 to 75 where spreads
are around zero or even negative. This finding matches the outcome
of our directional forecast metrics and suggests an overall lower
predictive power during the middle quarter-hours of the day. On the
other hand, its surrounding off-peak equivalents feature remarkably
high spreads. Some hours exhibit price differences around 2€/MWh.
Even under the assumption of negative peak spreads, the overall average
delta of more than 70Cent/MWh allows for the conclusion of economic
gains to be made in our case study. 
\begin{table*}
\centering{}\includegraphics[scale=0.68]{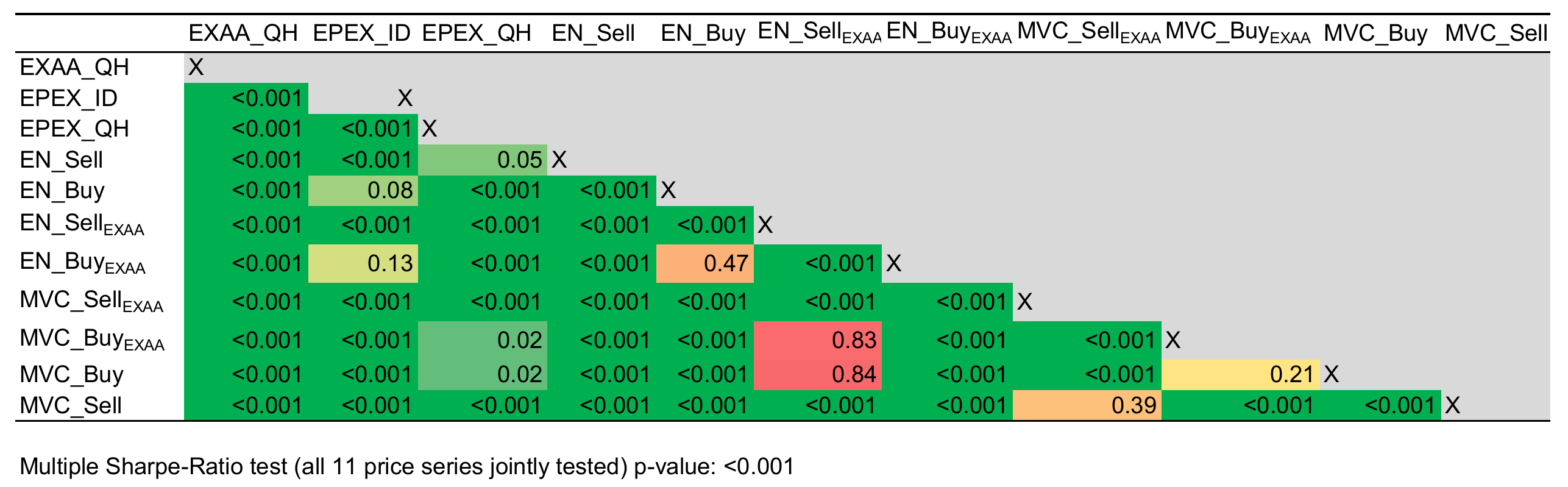}\caption{Test results reported as p-values for two Sharpe-Ratio equality tests.
The null hypothesis states that Sharpe-Ratios are equal and can be
rejected with sufficiently low p-values. We apply two different tests,
a pairwise test reported in the table and a test that jointly considers
all Sharpe-Ratios. }
\end{table*}
\\
\hspace*{0.5cm}Overall, we need to mention that a very primitive
strategy based on two point forecasts yields the most attractive economic
benefits albeit the test statistics before have revealed the limitations
of our point forecasts to binary prediction applications. The more
complex mean-variance optimization approach could not entirely live
up to the expectations. The strategy did not provide any spread benefits,
only a good Sharpe-Ratio and risk-averse portfolio structures. However,
the Markowitz optimization was the less volatile portfolio choice
with the lowest standard deviations. Despite the missing spread benefit,
its price level was exactly between the two individual exchanges and
marks the best alternative for risk averse market participants.\\
\hspace*{0.5cm}To be more concrete on numbers, we assume an equally
distributed 50MW QH spread position based on the Base\textsubscript{Sell\_EXAA}
and{\footnotesize{} }Base\textsubscript{Buy\_EXAA} forecasts. If
a market participant follows our EXAA base strategy from 07.10.2016
to 31.05.2018, savings of €325,080 for a buyer or additional revenues
in the same range for a seller are to be realized under the assumption
of no extra fees and access to VWAP prices.

\section{Conclusion and outlook }

\label{sec:outlook} We contributed to a blind spot in the current
literature by analyzing quarter-hourly German spot markets. The general
tendency towards more volatile power grids necessitated the introduction
of a quarter-hourly intraday call auction and the possibility to trade
quarter-hours in continuous intraday trading. Our paper provides the
first detailed discussion on how to forecast these markets ex-ante.
We have applied modern regression techniques, namely the elastic net
estimator that automatically penalizes features that do not add any
insight, and compared the outcome with classical linear regression
models. One of the peculiarities of German spot markets is the existence
of a variety of trading opportunities. In particular, the Austrian
EXAA offers a first day-ahead indication on quarter-hours that can
be delivered into the German grids. To account for that, we have applied
the EXAA as a standalone naive estimate as well as an input for our
more advanced regression models. We found that the intraday auction
is easier to predict compared to ongoing trading. Our EN-based prediction
method provides high forecasting accuracy and outperforms the considered
benchmark models. When we add the available EXAA prices, the results
are even more convincing. This assumption was further confirmed by
the popular Diebold-Mariano test that revealed a statistically sound
outperformance of all models, but EXAA ones and the EN one in particular,
over the naive benchmark. Surprisingly, this finding does not hold
true for the continuous intraday market. Our forecast models revealed
only minor increases in performance and fewer quarter-hours where
the Diebold-Mariano statistics suggest better results than the benchmark.
EXAA prices only mattered to a small extent. Another interesting aspect
occurred in the construction of input factors. We initially expected
the expert choice model to comprise all relevant factors, but the
outperformance of the full model group proved us wrong. When adding
every possible input, the OLS-based LM models ran into problems due
to the massive set of regressors but the elastic net and its feature
selection revealed lower error metrics. \\
\hspace*{0.5cm}If we recap the times of trading and forecasting,
a problem arises. The QH auction is estimated shortly after the data
has been published, i.e., uses the most current freely available inputs,
whereas the last hours of continuous trading are determined 24 hours
later. This situation could lead to new information. However, we have
neglected this last facet and have simultaneously predicted both markets
to evaluate the economic effects of our forecasts. Their standalone
information might help regulators or grid operators, but we deliberately
focus on a market player application and derive portfolio strategies
with both EXAA and non-EXAA-enriched estimations. We introduced a
straightforward ``sell in the high and buy in the low market'' rule
for the first set of portfolios and expanded the second group by a
Markowitz mean-variance approach. We were able to demonstrate that
the low/high strategies perform best, leading to considerable spreads
and attractive benefits for either a net buyer or a net seller. The
Markowitz approaches did not show any economic improvements in the
form of favorable spreads but delivered a maximum Sharpe-Ratio portfolio.
So even if market players seek to follow traditional mean-variance
strategies under the precept of risk-aversion, a precise quarter-hourly
forecast could deliver a suitable input for estimated returns. \\
\hspace*{0.5cm}At the same time, we must acknowledge that the basic
setup, despite its decent gains, was a rather simple one and could
be extended. We assumed a stable net buy or sell position in all QHs
and only roughly considered term-structure effects. A proper analysis
of weekends, peak/off-peak patterns or the aforementioned trading
and prediction time could yield beneficial insights. The same counts
for the point predictions itself. What if we continuously forecast
quarter-hourly prices once new information is published? Or how does
accuracy change if we add more accurate vendor data? We have just
focused on linear models in our study but of course there are other
non-linear prediction models such as random forests available. For
instance, a study in \citet{ludwig2015putting} has shown that lasso
estimators provided comparable results to random forests in EPEX day-ahead
predictions. But does this hold true for quarter-hourly markets as
well? Another point of possible criticism arises from the high/low
portfolio. The individual forecasts were combined to a directional
estimation. One could also discuss available directional forecast
approaches and simplify the forecasting problem to the binary one
that is utilized in the portfolio application. 

\section*{Acknowledgments}

The valuable contributions of anonymous referees are gratefully acknowledged.
This work was partially supported by the German Research Foundation (DFG, Germany) and the National Science Center (NCN, Poland) through BEETHOVEN project IMMORTAL (Investigating Market Microstructure and shOrt-term pRice forecasTing in intrA-day eLectricity markets) grant no. 379008354 (to FZ).

\renewcommand*{\bibfont}{\footnotesize}
\begingroup \raggedright \sloppy

\appendix

\section*{Appendix A. Supplementary data}

Supplementary data to this article can be found online at \href{https://data.mendeley.com/datasets/2trdgv8wrp/1}{DOI: 10.17632/2trdgv8wrp.3}.\bibliographystyle{model5-names}
\bibliography{Literature}

\endgroup
\end{document}